\documentclass[emulateapj]{aastex}
\usepackage{lscape}

\newcommand{\xray}{\mbox{X-ray}}
\def\simlt{\mathrel{\hbox{\rlap{\hbox{\lower4pt\hbox{$\sim$}}}\hbox{$<$}}}}
\def\simgt{\mathrel{\hbox{\rlap{\hbox{\lower4pt\hbox{$\sim$}}}\hbox{$>$}}}}

\def\CIVdblt{{\rm C~}\kern 0.1em{\sc iv}~$\lambda\lambda 1548, 1550$}
\def\MgIIdblt{{\rm Mg~}\kern 0.1em{\sc ii}~$\lambda\lambda 2796, 2803$}
\def\NVdblt{{\rm N}\kern 0.1em{\sc v}~$\lambda\lambda 1238, 1242$}  
\def\OVIdblt{{\rm O}\kern 0.1em{\sc vi}~$\lambda\lambda 1031, 1037$}
\def\SiIVdblt{{\rm Si~}\kern 0.1em{\sc iv}~$\lambda\lambda1394, 1403$}
\def\AlIIIdblt{{\rm Al~}\kern 0.1em{\sc iii}~$\lambda\lambda1855,1863$}
\def\FeIIdblt{{\rm Fe~}\kern 0.1em{\sc ii}~$\lambda\lambda 2383, 2600$}

\def\AlII{\hbox{{\rm Al~}\kern 0.1em{\sc ii}}}
\def\AlIII{\hbox{{\rm Al~}\kern 0.1em{\sc iii}}}
\def\CaI{\hbox{{\rm Ca}\kern 0.1em{\sc i}}}
\def\CaII{\hbox{{\rm Ca}\kern 0.1em{\sc ii}}}
\def\CrII{\hbox{{\rm Cr}\kern 0.1em{\sc ii}}}
\def\CII{\hbox{{\rm C~}\kern 0.1em{\sc ii}}}
\def\CIII{\hbox{{\rm C~}\kern 0.1em{\sc iii}}}
\def\CIV{\hbox{{\rm C~}\kern 0.1em{\sc iv}}}
\def\CV{\hbox{{\rm C}\kern 0.1em{\sc v}}}

\def\HI{\hbox{{\rm H~}\kern 0.1em{\sc i}}}
\def\HII{\hbox{{\rm H~}\kern 0.1em{\sc ii}}}
\def\Lya{\hbox{{\rm Ly}\kern 0.1em$\alpha$}}
\def\Lyb{\hbox{{\rm Ly}\kern 0.1em$\beta$}}
\def\Lyg{\hbox{{\rm Ly}\kern 0.1em$\gamma$}}
\def\Lyfive{\hbox{{\rm Ly}\kern 0.1em$5$}}
\def\Lysix{\hbox{{\rm Ly}\kern 0.1em$6$}}
\def\Lyseven{\hbox{{\rm Ly}\kern 0.1em$7$}}
\def\Lyeight{\hbox{{\rm Ly}\kern 0.1em$8$}}
\def\Lynine{\hbox{{\rm Ly}\kern 0.1em$9$}}
\def\Lyten{\hbox{{\rm Ly}\kern 0.1em$10$}}
\def\HeI{\hbox{{\rm He}\kern 0.1em{\sc i}}}
\def\HeII{\hbox{{\rm He}\kern 0.1em{\sc ii}}}
\def\FeI{\hbox{{\rm Fe~}\kern 0.1em{\sc i}}}
\def\FeII{\hbox{{\rm Fe~}\kern 0.1em{\sc ii}}}
\def\FeIII{\hbox{{\rm Fe~}\kern 0.1em{\sc iii}}}
\def\MnII{\hbox{{\rm Mn}\kern 0.1em{\sc ii}}}
\def\MgI{\hbox{{\rm Mg~}\kern 0.1em{\sc i}}}
\def\MgII{\hbox{{\rm Mg~}\kern 0.1em{\sc ii}}}
\def\MgIII{\hbox{{\rm Mg~}\kern 0.1em{\sc iii}}}
\def\MgIV{\hbox{{\rm Mg~}\kern 0.1em{\sc iv}}}
\def\NaI{\hbox{{\rm Na}\kern 0.1em{\sc i}}}
\def\NV{\hbox{{\rm N}\kern 0.1em{\sc v}}}
\def\NII{\hbox{{\rm N}\kern 0.1em{\sc ii}}}
\def\NIII{\hbox{{\rm N}\kern 0.1em{\sc iii}}}
\def\NiII{\hbox{{\rm Ni~}\kern 0.1em{\sc ii}}}
\def\OVI{\hbox{{\rm O}\kern 0.1em{\sc vi}}}
\def\OI{\hbox{{\rm O}\kern 0.1em{\sc i}}}
\def\OII{\hbox{[{\rm O}\kern 0.1em{\sc ii}]}}
\def\SiII{\hbox{{\rm Si~}\kern 0.1em{\sc ii}}}
\def\SiIII{\hbox{{\rm Si~}\kern 0.1em{\sc iii}}}
\def\SiIV{\hbox{{\rm Si~}\kern 0.1em{\sc iv}}}
\def\SII{\hbox{{\rm S}\kern 0.1em{\sc ii}}}
\def\SIII{\hbox{{\rm S}\kern 0.1em{\sc iii}}}
\def\SIV{\hbox{{\rm S}\kern 0.1em{\sc iv}}}
\def\TiII{\hbox{{\rm Ti}\kern 0.1em{\sc ii}}}
\def\ZnII{\hbox{{\rm Zn~}\kern 0.1em{\sc ii}}}
\newcommand{\kms}{\hbox{km~s$^{-1}$}}

\def\kms{\hbox{km~s$^{-1}$}}

\newcommand{\noprint}[1]{}
\newcommand{\figsetstart}{{\bf Fig. Set} }
\newcommand{\figsetend}{}
\newcommand{\figsetgrpstart}{}
\newcommand{\figsetgrpend}{}
\newcommand{\figsetnum}[1]{{\bf #1.}}
\newcommand{\figsettitle}[1]{ {\bf #1} }
\newcommand{\figsetgrpnum}[1]{\noprint{#1}}
\newcommand{\figsetgrptitle}[1]{\noprint{#1}}
\newcommand{\figsetplot}[1]{\noprint{#1}}
\newcommand{\figsetgrpnote}[1]{\noprint{#1}}

\begin{document}

\title{Testing the Possible Intrinsic Origin of the Excess Very Strong MgII Absorbers Along GRB Lines-of-Sight
}

\shorttitle{Possible Origin of MgII Absorbers}
\author{A. Cucchiara\altaffilmark{1},  T. Jones\altaffilmark{1}, J.~C. Charlton\altaffilmark{1}, D. B. Fox\altaffilmark{1}, D. Einsig\altaffilmark{1} and A. Narayanan\altaffilmark{2}}

\email{cucchiara@astro.psu.edu, tjones@astro.psu.edu, dfox@astro.psu.edu, charlton@astro.psu.edu, deinsig@astro.psu.edu, anand@astro.wisc.edu}

\altaffiltext{1}{Department of Astronomy \& Astrophysics, 525 Davey Lab., Pennsylvania State
University, University Park, PA 16802, USA}
\altaffiltext{2}{Department of Astronomy, University of Wisconsin-Madison, 475 North Charter Street, Madison, WI 53706}

\begin{abstract}
The startling discovery of \cite{ppc06} that the
frequency of very strong ($W_r(2796)>1$\,\AA) {\MgII} absorbers along
gamma-ray burst (GRB) lines of sight ($[dN/dz]_{\rm{GRB}} = 0.90$) is
more than three times the frequency along quasar lines of sight
($[dN/dz]_{\rm{QSO}} = 0.24$), over similar redshift ranges, has yet to
be understood.  In particular, explanations appealing to dust
anti-bias in quasar samples, partial covering of the quasar sources,
and gravitational-lensing amplification of the GRBs have all been
carefully examined and found wanting.  We therefore reconsider the
possibility that the excess of very strong {\MgII} absorbers toward
GRBs is intrinsic either to the GRBs themselves or to their immediate
environment, and associated with bulk outflows with velocities as
large as $v_{\rm max}\sim 0.3c$.  In order to examine this
hypothesis, we accumulate a sample of 27 $W_r(2796) >
1$\,\AA\ absorption systems found toward 81 quasars, and compare their
properties to those of 8 $W_r(2796)>1$\,\AA\ absorption systems found
toward 6 GRBs; all systems have been observed at high spectral
resolution ($R = 45,000$) using the Ultraviolet and Visual Echelle
Spectrograph on the Very Large Telescope.  We make multiple
comparisons of the absorber properties across the two populations,
testing for differences in metallicity, ionization state, abundance
patterns, dust abundance, kinematics, and phase structure.  We find no
significant differences between the two absorber populations using any
of these metrics, implying that, if the excess of absorbers along GRB
lines of sight are indeed intrinsic, they must be produced by a
process which has strong similarities to the processes yielding strong
{\MgII} systems associated with intervening galaxies.  Although
this may seem a priori unlikely, given the high outflow velocities
required for any intrinsic model, we note that the same conclusion was
reached, recently, with respect to the narrow absorption line
systems seen in some quasars.

\end{abstract}


\section{Introduction and Motivation for This Study}
\label{sec:intro}
Intervening metal absorption line systems have been routinely observed
along the lines of sight to cosmological gamma-ray bursts (GRB) since
the first optical/UV spectrum of a GRB afterglow was obtained in 1997
\citep{mdk+97}. In particular, {\MgIIdblt} resonance doublet
absorption, observable from the ground over redshifts $0.4 \lesssim z
\lesssim 2.2 $, has been detected in many GRB afterglow spectra, at
redshifts well separated from that of the highest-redshift absorber in
the system, which is typically associated with the GRB host galaxy.

Strong intervening {\MgII} absorbers are a familiar subject of quasar
absorption-line studies, and have been used for decades in studies of
intermediate-redshift galaxies and their environments.  Indeed, it is
now known that the presence of absorption with {\MgII} $\lambda$2796
rest frame equivalent width $W_r(2796) > 0.3$~{\AA} (so called ``strong" absorption systems)
is commonly ($\sim 75$\%
of all cases) associated with the presence of a nearby (within 60~kpc
projected distance) $\sim$ 0.1--5 \,$L_\star$ galaxy along the
line-of-sight \citep{kcs08}.

The detailed physical picture of the strong \MgII\ absorbers continues
to be elaborated.  

Among the strongest ($W_r(2796)>1$~{\AA}) {\MgII} absorbers at
$z<1.65$, approximately 80\% are also damped Lyman-alpha systems
(DLAs; \citealt{rtn06}).
Imaging of the quasar fields for a subset of even-stronger
($W_r(2796)>2.7$~{\AA}) {\MgII} absorbers at low redshift ($0.42 < z <
0.84$) indicates that interactions, pairs, and starburst related
phenomena are likely to be responsible \citep{bmp207}.
In roughly the same redshift range, \citet{bmp206} find an anti-correlation
between {\MgII} equivalent width for each system and the mass of its
associated dark matter halo based on halo masses predicted from clustering analysis.  Since the {\MgII} equivalent
width of strong absorbers is determined primarily by the velocity
dispersion of its constituent absorbing clouds, this relationship is
consistent with a starburst picture for the strongest {\MgII} systems,
but not with structure within individual virialized haloes.
Finally, \citet{ppc06} concluded, over a larger redshift range ($0.35
< z < 2.7$), based on the kinematics of $W_r(2796)>1$~{\AA} absorbers,
that these structures are related to superwinds, rather than to
large-scale gas infalls in galaxy halos.
Indeed, \citet{ntr05} find a rapid decline in the incidence of
$W_r(2796)>2$~{\AA} systems with decreasing redshift, consistent with
the expectations from superwinds since these are thought to 
increase in concert with the global star formation rate.

GRBs and their afterglows are promising probes of the high-redshift
Universe due to their brightness and observed redshift distribution,
now extending beyond $z=6.7$.  High-resolution spectroscopy of the
brightest afterglows has been used to study GRB host galaxies, and the
subgalactic environs of GRBs, up to $z=6$. With
respect to intervening absorption systems, one would naively expect
GRB lines of sight to be equivalent to quasar lines of sight, including strong {\MgII}
absorbers.  However, for very strong ($W_r(2796) > 1$~\AA) {\MgII}
absorbers, a puzzling difference in the redshift path density,
\emph{dN/dz}, between GRB and QSO sightlines has been discovered.  The
frequency of these absorbers along GRB lines of sight is more than
three times larger ($[dN/dz]_{\rm{GRB}} = 0.90$) than the frequency
along quasar sightlines ($[dN/dz]_{\rm{QSO}} = 0.24$) covering a similar
range of redshifts \citep{ppc06}.

Since the discovery of this factor of $\approx$ 3 discrepancy, several
hypotheses have been advanced to explain it:  
(1) The number of {\MgII} systems along quasar lines of sight has been
suppressed, due to a bias within quasar samples towards brighter
objects lacking, e.g., dusty foreground systems along the line of
sight; 
(2) The relative beam sizes of quasar and GRB afterglow emitting
regions lead to partial covering in quasar spectra, and increased
numbers of strong absorption systems in GRB spectra; 
(3) Gravitational lensing by the mass concentrations associated with
strong absorption systems magnifies the GRB and its afterglow,
increasing the probability of detection and spectroscopic observation;
(4) A dominant number of the strong absorption systems in GRB spectra
are physically associated with the GRB environment, the GRB itself, or
both.

We will now briefly review the status of each of these hypotheses as a
way of motivating the present study, which focuses on the fourth
hypothesis, that the excess of very strong {\MgII} absorption
systems in GRB afterglow spectra is intrinsic to the GRB or its
environment.  For any of the other hypotheses to provide an
explanation of the full effect, the resulting biases would have to be
quite large.  The intrinsic hypothesis, by contrast, is required to
produce on average roughly one intrinsic, high-velocity absorption system per
GRB.

\subsection{Bias Due to Dust}
\label{sec:introdust}

If the very strongest intervening {\MgII} absorbers tended to arise in the
dustiest environments they would diminish the observed magnitude of
a background quasar.  In such a way, optical magnitude-limited
quasar absorption-line surveys might be biased toward the brighter 
quasars that do not have the strongest {\MgII} systems in the 
foreground. Although gamma ray bursts would suffer the same dust bias 
effect, this effect is less important in determining whether a high
signal-to-noise, high-resolution spectrum can be obtained than is the
speed at which follow-up observations are possible.  Thus we
would expect that gamma ray bursts that have very strong, dusty
{\MgII} absorbers in the foreground would still be present in
a sample of high-resolution GRB optical spectra.  GRB spectra
should therefore have more very strong {\MgII} absorbers on average
than do quasar spectra.

The main problem with this explanation for the excess of 
$W_r(2796)>1$ {\AA} absorbers in GRB spectra is that the magnitude of
the effect is not nearly large enough to explain a factor of more than 
three excess. \citet{mnt08} found that $\lesssim 2$~\% of quasars are 
absent from optical surveys due to reddening and extinction from
strong {\MgII} systems ($1 < W_r(2796) < 3$~{\AA}).  Even for
$W_r(2796) > 5$~{\AA} systems, which are too rare to affect enough 
quasars to explain the observed excess, slightly less than
half of the quasars with such foreground absorbers would be missed
in a magnitude-limited sample.
More directly, \citet{evl06} estimated the extinction due to a
$W_r(2796)=1.87$ ~{\AA}
intervening system along the line of sight of GRB\,060418 at redshift $z = 1.106$. The amount
of extinction derived is $E({\rm B}-{\rm V}) = 0.08$~mag, which is not
sufficient to obscure sufficient quasars to explain the observed discrepancy
in $dN/dz$.

\subsection{Partial Beam Coverage Explanations}

An interesting explanation has been proposed by \citet{fbs07},
assuming that GRB emitting regions are generally smaller than those
of quasars.  In order to avoid obvious signatures of partial
covering in the case of quasar strong {\MgII} absorbers, \citet{fbs07}
derived a cloud density profile with a constant density core
surrounded by a power-law density profile.  The density profile
sampled by the beam depends on the size of the beam
relative to the core size of the absorber and on the impact
parameter. Small beams that pass through only the center of the cloud
sample much higher column densities than beams larger than the core
for which high optical depth absorption is diluted.  Because the GRB beam size changes in time,
one prediction
of this scenario  is  variability
in the strength and the structure of their absorption lines.
Such variability has not yet been seen \citep{twl07}, although it was
once suggested \citep{hsd07} in the case of GRB 060206.
More importantly, \citet{phc07} point out that signatures of partial
covering should still be seen in high-resolution spectra unless the
cloud density profiles are fine-tuned to an unreasonable degree.
They also show that it is unlikely that an excess of the needed
magnitude could arise from this effect.
Finally, there is no way to set up the clouds producing the numerous
distinct absorption components seen in strong {\MgII} absorbers such that
the line of sight passes through the inner high-density
core of {\it all} of the clouds  \citep{pvs07}.

\subsection{Lensing Amplification of GRB Beam-size}

Following an analogy with BL~Lac objects as studied by \citet{sr97},
the idea that a GRB line of sight intersects more absorbers because
the emitting region is microlensed by a gravitational potential has
also been considered (e.g. \cite{lp298, gls00, bh05, huY06}).  In
particular, considering binary lensing and \emph{double} magnification
bias, \citet{bh05} estimate that there is a 60\% chance that a given
GRB is microlensed. On the other hand, using a sample from Sloan
Digital Sky Survey (SDSS) quasars, the magnification factor has been
calculated to be $\mu \lesssim 1.10$ \citep{mnt08}. This value is not
sufficient to lead to the observed excess of very strong {\MgII}
absorbers towards GRBs.  Furthermore, if the lensing effect is
significant for GRBs, we would expect strong lensing in some cases,
along with associated multiple images, ``repeating'' GRBs, and
``bumps'' in GRB optical light curves \citep{lp298}, none of which are
commonly seen.

At the same time, it is important to consider whether a significant
magnification bias could result from an underlying source luminosity
distribution (i.e.\ $\log N$--$\log L$) that is steep at its faint
end. \citet{ngg06,ngg08} have argued that this is the case, with
power-law index $\alpha \gtrsim -3$; 
however, a statistical sample will be needed to test this
hypothesis which is based on energetic arguments.  \citet{zd08} found
that using an internal shock model the faint end slope of the GRB
optical luminosity function can be fitted by a power law index with
$\alpha \approx -0.6$ , as also noticed by \cite{pvs07}, which would
be too flat to explain the discrepancy.


\subsection{An Origin Intrinsic to GRBs or Their Environments}
\label{sec:introintrin}

In view of the dramatic excess of $W_r(2796)>1$~{\AA} absorbers toward
GRBs and the problems with the explanations discussed above, we
should consider further the idea that the GRBs themselves or their
immediate environments are responsible for material observed
as {\MgII} absorption.  Since the observed excess is apparent
over a substantial redshift range, the material would have to be
accelerated to at least $0.2$--$0.3c$ relative to the GRB host
galaxy.  At first glance it might seem implausible that such
high velocity ejection could be consistent with such narrow
lines as are observed as components in the strong {\MgII}
absorbers \citep{ppc06}.  However, it is important to note that
similar high velocity narrow absorption lines  are known to
be ejected in the accretion disk winds of quasars \citep{mce07,gb+08,rhn07,nhb04, nmc07}.  
This phenomenon also has yet to be explained by
accretion disk wind models. Clearly it is possible for dense, coherent
clouds of gas to hold together, even in mildly-relativistic
outflows, at least in this case.  So it seems possible
that the same could be the case for GRBs.

We know that the GRB forward shock reaches distances of $d\gtrsim
10^{14}$\,cm from the central engine, after which it is decelerated
by the ambient medium. Over this distance it interacts
with the circumstellar (and interstellar) medium, generating a
relativistic shock that produces the afterglow emission via
synchrotron radiation of accelerated electrons.  The
highly-relativistic motion of this shock, expected to have a Lorentz
factor of $\Gamma\gtrsim 300$ at early times and evolving according to
the Blandford-McKee solution (\citealt{bmc76}), sets severe constraints on
the properties of any burst-related gas that might be responsible for
absorption features in the afterglow spectrum.

Beyond the region affected by the physical shock, GRB high-energy
photons are able to ionize gas within $d\simlt 10$\,pc from the
central engine, as demonstrated by \xray\ spectroscopy of GRB\,050904,
which exhibited a time-variable absorbing column \citep{clr07,whf07}.

At distances greater than $100$~pc, there are signatures of absorption
by neutral hydrogen, dust, and metal enriched and molecular gas
\citep{psp09}.  UV pumping of atomic states by the burst prompt and
early afterglow emission is estimated to occur at distances of $d\sim
50$\,pc (for {\MgI} absorption in GRB\,050111; \citealt{pcb+06}) to
1.7~kpc (for \NiII\ and \FeII\ absorption in GRB\,060418;
\citealt{vls07}).

If the GRBs emission photoionizes {\MgII} clouds out to $d\gtrsim
100$\,pc, this should lead to an increase in temperature to $T \gg
10^4$\,K.  It would thus be necessary to rely on the existence of a
dense, metal-rich cloud which is not penetrated by the relativistic
shock and which cools faster than the surrounding material.  As
proposed by \citet{vlg07}, the progenitor's Wolf-Rayet wind is capable
of providing dense material at the required distance.  Unfortunately,
the likely velocities are an order of magnitude less than is needed to
explain the velocities of the GRB excess absorbers considered here.

The need to supply the near vicinity of GRBs with a copious amount of
iron-enriched, fast-moving gas provided one of the initial motivations
for the ``supranova'' model \citep{wdl02,kg02,vgl01}, which proposed
that a supernova mere weeks to months prior to the GRB was responsible
for dispersing this gas, and that the collapse of the metastable,
fast-spinning neutron star born in the supernova later generated the
burst itself.  The original observational stimulus to these models was
provided by claims of detection of high equivalent-width iron emission
lines in afterglow \xray\ spectra \citep[][e.g.,]{pgg00}; these claims
are no longer favored \citep{shr05}.  However, given that the model
produces fast-moving gas with speeds of $v\simlt 0.3c$ it may be worth
mentioning again in the present context.

Even if it is necessary for a supernova to explode prior to the GRB,
to produce fast-moving gas along the line of sight, it may not be
necessary for the two to be directly related.  \cite{wbh03} suggest
that GRB\,021004 was located in a region where the ISM was metal-rich
due to high velocity ejecta from a hydrogen-rich ``supernova'' that
exploded a few months before the GRB.  In this particular case, the
detected Lyman-$\alpha$ components also indicate absorption by a
clumpy medium. This material has been accelerated then by the GRB
blast wave (see \cite{wdl02} for a detailed analysis).  In the
limiting case where the absorbing gas must be relativistic, $v\sim
0.3c$, it may be necessary to invoke a ``hypernova'' (SN\,1998bw-like
SN) rather than a more ordinary type of SN \citep{mdm02,mvd08,sbp08}.

Ultimately, the required apparent velocities needed to explain the
excess GRB absorbers are not unusual for quasar outflows.  Intrinsic
narrow absorption lines, {\CIV} and {\SiIV}, have been observed in
quasar spectra \citep{mce07,gb+08,rhn07, nhb04, nmc07} showing
blueshifts consistent with an ejection velocity in the range
5,000--70,000 $\rm{km} ~\rm{s}^{-1}$.  A comprehensive model to
understand these intrinsic feature has still to be developed, but the
similarities with the GRB case are intriguing.

Whether or not these ideas for fast-moving absorption systems are
reasonable, it is necessary in addition to understand how they fit
into the larger context of the GRB immediate environs and host galaxy,
and their associated absorption features.  The bulk of the host galaxy
absorption is apparently coming from $>1$~kpc, that is, from the host
galaxy at large rather than the immediate GRB environment; yet it is
clearly affected by the GRB radiation field, at least in those cases
where metastable lines are observed.  These systems are not counted in
the excess of very strong {\MgII} absorbers found by \cite{ppc06}, but
it is important to consider how the absorption signature from
high-velocity material more local to the GRB might compare to that of
the host galaxy.

\subsection{Motivation for This Study}
\label{sec:motivation}

We aim to consider the similarities and differences between the
absorption profiles of: (1) very strong {\MgII} absorbers found in
quasar spectra; (2) very strong {\MgII} absorbers found in GRB
spectra, which are a combination of the same objects found in quasar
spectra and a separate population yet to be understood; and (3) the
GRB host galaxy absorption.  Naively we might expect that the
absorbers responsible for the factor of three excess of very strong
{\MgII} absorbers along GRB sightlines should differ in some way from
the usual quasar absorption line systems.  We thus compare quasar
absorption line systems taken from the VLT/UVES archive with systems
seen in absorption in GRB spectra obtained with the same instrument
and configuration.  We consider the kinematics of the {\MgII}
profiles; the ratios of {\MgII} equivalent widths to those of other
transitions, including dust tracers; the relationship between high and
low ionization transitions; and the possible presence of metastable
lines.

In the next section, we describe the VLT/UVES datasets and our
analysis methods.  In \S\ref{sec:results} we present the results of
our comparisons between the very strong {\MgII} absorbers seen toward
quasars and GRBs.  In \S\ref{sec:discussion} we discuss the
implications of our result that there are no apparent differences
between these populations, in particular, considering the implications
for the hypothesis of an intrinsic origin for a majority of the GRB
absorbers.  \S\ref{sec:summary} summarizes our conclusions.

\section{Dataset and analysis}
\label{sec:data}

Our GRB dataset consists of 6 GRBs observed with the Ultraviolet and
Visible Echelle Spectrograph (UVES) mounted on VLT. These include all
high resolution UVES spectra that were accessible to the public before
August 2008.  The details of the observations, including the time that the
observations were obtained, the wavelength coverage, and the
exposure time, are listed in Table~1. In the last column of the table 
we report previous works on the GRBs host galaxy.

The quasar dataset included 81 QSOs, also obtained with UVES/VLT.  These
quasars, listed in Table~1 of \cite{nmc07}, are those for which
data were available before June 2006.  All observations of a given
quasar were combined with $S/N$ weighting, after scaling by the ratio of
the median ratio of the number of counts in the best exposure to the counts
in the given exposure.

The GRB and quasars spectra were reduced  using 
standard UVES/VLT tools under the \emph{MIDAS} environment.
Because of possible variability, the different exposures of the
same GRB were not combined.   A conversion to a heliocentric
vacuum scale was applied to the final spectra.  Continuum
fitting and normalization was performed using the IRAF
SFIT procedure\footnote{IRAF is distributed by the National
Optical Astronomy Observatories, which are operated by AURA,
Inc., under cooperative agreement with NSF}, by dividing the
spectrum into 3000 pixels segments and fitting each segment
separately.  The spectra were normalized by the resulting
continuum fit.

The signal-to-noise ratio ratio (S/N) of the QSO spectra ranges from
$\sim20-100$ per pixel over most of the wavelength coverage.  The GRB
spectra are somewhat noisier due to the limitations caused by the need
for rapid followup before the GRB fades.  We searched both the QSO and
GRB spectra for {\MgII} doublets in regions redward of the {\Lya}
forest.  Formally, we applied a $5\sigma$ detection limit to
absorption 
features. We first assumed that every detection represents the \MgII ~$\lambda$2796 
component of the {\MgII} doublet. A candidate {\MgII} system was considered if there was at least a 2.5 $\sigma$
detection of the corresponding $\lambda2803$ line for the same redshift.
We also compared the two line profiles, which should show similar subcomponents, and 
then we calculated the equivalent width of the blue component via Voigt-profile fitting.

Since our
focus is on very strong {\MgII} absorbers, with equivalent width
$W(\lambda2796)>1$ \AA, we are easily detecting all systems of
interest.  For detected systems we then searched the expected
locations of other ions, including {\Lya}, {\Lyb},
{\MgI} $\lambda 2853$,  {\FeII} $\lambda$2374, 2383, 2587,
and 2600, {\SiII} $\lambda$1260, {\CII} $\lambda$1335,
{\SiIVdblt}, {\CIVdblt} (the complete list can be found in Tables 3,
4, and 5).  

Several of these transitions
are only covered for the very highest redshifts in our sample, and
thus could not be used for a statistical comparison.
We also considered dust tracers such as {\ZnII} $\lambda$2026,2063,
{\CrII} $\lambda$2056,2062,2066, {\NiII} $\lambda$1710,1752, and
{\MnII} $\lambda$2577,2594,2606.  Whenever these transitions
are covered we measured the equivalent widths or $3\sigma$ equivalent-width 
upper limits.  When blends with transitions from systems at other
redshifts were identified we considered the measured equivalent width
as an upper limit.

\section{Results}
\label{sec:results}

In our search of the 6 GRB spectra listed in Table 1, we found 8 $W_r(2796) > 1$~{\AA}
{\MgII} doublets.  The redshift path length for our GRB search was 9.9, giving a
$dN/dz = 0.8\pm 0.3$, consistent with the value from \citet{ppc06}, obtained from 14 GRBs spectra.  Similarly,
we identified $27$ $W_r(2796) > 1$~{\AA} {\MgII} doublets over a redshift
pathlength of 77.3 towards 81 QSOs.  We derive $dN/dz = 0.35\pm 0.07$ for very strong
{\MgII} absorbers observed toward quasars, which is also consistent with the
much larger Sloan sample of \citet{ntr05}. 
Some GRBs of our sample were already present in \citep{ppc06}, so this consistency is not 
completely surprising. 
System plots for the GRB absorbers, including
various transitions that provide useful constraints, are shown in velocity space in Fig.~\ref{fig:fig1a}-h.  Similar plots for the quasar absorbers are given in Fig.~\ref{fig:fig2a}-aa.
Basic information about the absorbers, for both GRBs and quasars, is given in Table~2.
Rest frame equivalent widths of selected transitions that were detected in one or more systems
are given in Tables~3, 4, and 5, along with equivalent width limits in cases where the transition
was covered but not detected.

The excess of strong {\MgII} absorbers along GRB lines of sight, as compared
to quasars, applies only for $W_r(2796) > 1$~{\AA}, though there appears to
be no greater an effect for even stronger systems \citep{ppc06}.  For our
VLT/UVES samples we plot $W_r(2796)$ as a function of redshift in Fig.~\ref{fig:fig3}d.
The equivalent width distributions of the GRB and quasar {\MgII} absorbers
were compared (including only those with $W_r(2796) > 1$~{\AA}) using a Kolmogorov-Smirnov
test (K-S test) and it was found that they are consistent with being drawn from the same distribution ($p=12\%$).  

It is also important to consider whether the redshift
distributions of the two samples are the same, since $W_r(2796)>2$~{\AA}
{\MgII} absorbers are known to evolve in the sense that they are less common
at low redshift.  A redshift difference could therefore lead to a difference
in the equivalent width distributions.  We find, however, that a K-S test
shows that the redshifts of the GRB and quasar {\MgII} absorber
samples are consistent of being drawn from the same distribution ($p=30\%$).

A difference in the redshift distributions between the two samples
could also be indicative of a concentration of GRB {\MgII} absorbers
at lesser ejection velocities.  We therefore also show in
Fig.~\ref{fig:fig3}b the cumulative distributions of relative
velocities for the two samples, normalized to 1.  We convert the
quasar {\MgII} absorber redshifts to relative velocities merely to
facilitate the comparison.  There is no significant difference between
these distributions, though we note that only one of the GRB absorbers
has an apparent ejection velocity greater than $0.4c$.

In this section we describe various comparisons of the kinematics and
the absorption in numerous chemical transitions for the 8 very strong
{\MgII} absorbers seen towards GRBs and the 27 seen towards quasars.
These detailed comparisons are facilitated by the high resolution of
the UVES/VLT spectra.  Fig.~\ref{fig:fig4} shows the absorption profiles of the
{\MgII} $\lambda$2796 line for the 8 GRB {\MgII} systems in the left panel and
a selected sample of 8 quasar systems at similar redshift in the right panel.  At face value, the
profiles are similar in terms of their equivalent widths and kinematic
structures. We will now examine this quantitatively, considering also
other elements and ions and their relation to the {\MgII} absorption properties.
We also note that there is no evidence for partial covering for the {\MgII} absorbers,
neither for the quasar or GRB cases.  Most lines/components are saturated 
over a finite extent in wavelength.

\subsection{{\MgII} Kinematics}
\label{sec:kinematics}

We make quantitative comparisons of the {\MgII} $\lambda$2796 profiles
of the quasar and GRB absorbers using some of the same statistics that
\citet{mcl07} used to consider evolution of strong {\MgII} absorbers.
In addition to the equivalent widths, we describe the {\MgII}
$\lambda$2796 absorption profiles by several kinematic indicators,
namely, the full velocity range, $\Delta V$, kinematic spread,
$\omega_v$, and $D$-index, $D$.  The full velocity range of a system
$\Delta V$ is the difference between the minimum and maximum
velocities at which absorption is detected while the kinematic spread,
as defined in detail in Appendix A of \citet{cv01},  is the optical depth weighted second moment of the
velocity.  This kinematic indicator is particularly sensitive to weak
components at high velocities from the central absorption.  Finally,
we estimated the $D$-index \citep{ell06,ell09}, defined as $D = 1000
\times (W_r(2796)[$\AA$])/ \Delta V [{\kms}]$, in order to indicate a
distinction between DLAs and lower column-density absorbers.  $D$
gives an indication of the fraction of the profile that has saturated
absorption in {\MgII} $\lambda$2796.

Figure~\ref{fig:fig5} presents the three kinematic indicators, $D$, $\omega_v$,
and $\Delta V$, for quasar and GRB {\MgII} $\lambda$2796 profiles as a function
of redshift.  There is no evolution apparent in $\omega_v$ or $\Delta V$ for
the quasar population.  There is a suggestion of an increase in $D$ with
decreasing $z$, but it is only significant at the 4\% level according to the Kendall
$\tau$ rank order test  \citep{fgh}.  If this trend is real it would
indicate that low-redshift, very-strong {\MgII} absorbers are more likely to have black
absorption across their profiles.  
Figure~\ref{fig:fig5} also shows the dependence of $D$ on $W_r(2796)$.
There is no significant correlation between these two quantities.

The main purpose of Figure~\ref{fig:fig5}, for the purpose of this study, is to compare the
quasar and GRB {\MgII} $\lambda$2796 profiles.  We found no significant
differences between the distributions of $D$, $\omega_v$, or $\Delta V$
between the two populations as evaluated using a Kolmogorov-Smirnov (K-S) test.
In particular, a K-S test comparing the \emph{D}-values between GRBs and QSOs 
are consistent of being drawn from the same distribution ($p=13.5\%$).

\subsection{Equivalent Width Comparisons}

Fig.~\ref{fig:fig6} plots the rest-frame equivalent widths of selected transitions
versus those of {\MgII} $\lambda$2796.  These particular transitions were selected for display because
they were accessible to analysis in our spectra and/or because they represent
important tracers of the physical conditions in the gas.  Rest frame equivalent widths
and equivalent width limits for other transitions are given in Tables 3, 4 and 5.
The basic result from Fig.~\ref{fig:fig6} is that there are no significant differences
between the GRB and quasar samples.  The {\MgI} $\lambda$2853, {\FeII} $\lambda$2600,
and {\FeII} $\lambda$2374 transitions are almost always covered and detected, but
the values for GRB absorbers span the range of values for quasar absorbers of similar $W_r(2796)$.  The detection of {\MgI} absorption implies that the absorbing
gas cannot be located within 50~pc of the GRB afterglow \citep{pcb+06}.
We therefore note that several of the GRB absorbers are among the lowest {\MgI}
equivalent widths for our sample, though there is no statistically significant
difference in the overall distributions.

The dust tracer  {\MnII} $\lambda$2577 is covered in many cases, though
often   not detected.  Again, the GRB and quasar absorbers have a similar
distribution in Fig.~\ref{fig:fig6}. Dust will be discussed further in \S~\ref{sec:dust}.
The higher-ionization line {\AlIII} $\lambda$1855 is only
covered for the highest-redshift quasars and GRBs but it can be noted that the
few GRB absorbers do not deviate significantly from the quasar absorbers in their
high-ionization content.  Details of the high-ionization kinematics, as traced by CIV, will be
discussed in \S~\ref{sec:highion}.

\subsection{{\FeII} to {\MgII} Ratio}
The ratio of {\FeII} to {\MgII} provides a good measure of the ionization parameters for
$\log U$ values greater than -4.0, where $U$ is the ionization potential for the particular transition.  Also, if the Fe/Mg ratio is small it could suggest
that the gas is $\alpha$-enhanced, while if it is large it is clear that Type Ia supernovae
have played a role in enriching the gas.  Alternatively, a lower {\FeII} to {\MgII}
ratio can arise in lower density gas, with a higher ionization parameter.
From Fig.~\ref{fig:fig7} we see, for the QSOs in particular, an evolution in the equivalent width ratio of
{\FeII} to {\MgII}, with an absence of small values at $z<1.2$ \citep{wcc09}.
We interpret this as an absence of $\alpha$-enhanced very-strong {\MgII} absorbers
at low redshifts due, presumably, to a diminishment of contributions of superwinds
to this class of absorber at $z<1.2$.  

There are not enough very strong {\MgII} absorbers found toward GRBs
to confirm whether the same trend in evolution of the {\FeII} to
{\MgII} ratio holds for the GRB absorbers.  However, there are two
GRBs with among the lowest {\FeII} to {\MgII} equivalent width ratios
for low redshift absorbers (the GRB\,060418 system at $z=0.655$ and
the GRB\,050820 system at $z= 0.691$).  If this tendency were
supported by a larger dataset, it would suggest that the low-redshift
GRB absorbers do indeed differ from the quasar population supporting
an origin at high redshift in the GRB environment.  However, we do not
have sufficient GRB data at this time to support this conclusion.

\subsection{Dust}
\label{sec:dust}

To estimate the dust depletion we used the ratio Fe/Zn.  This
quantity is a good tracer of depletion due to the diference in these
elements' refractory properties.  As shown in \citet{sav06}, dust
depletion tends to be higher for a higher {\ZnII} column density.  
Specifically, we used the equivalent width ratio of
{\FeII}~$\lambda$2374 to {\ZnII}~$\lambda$2026 as a dust indicator,
since we could directly measure this from our data.  Although a column
density ratio would be more physically meaningful, our goal is only to
make a comparison between the GRB and quasar absorbers, so we have
chosen to use the direct observable.

Based on the derived values of, for exampple, $W_r(2374)/W_r(2026)$, 
as shown in Fig.~\ref{fig:fig8}, we find again that the quasar and GRB populations
are similar.  However, the lowest data point, arising from the $z=1.106$ system
toward GRB 060418, deserves further comment.  We confirm the finding of \cite{evl06}
that this absorber is particularly dusty.  However, as mentioned in \S~\ref{sec:introdust},
this amount of dust in quasar {\MgII} absorbers would not be sufficient to prevent a typical quasar from
being observed in large-area surveys.  It might be an indication of a particularly
unusual environment in that GRB absorber.

\subsection{{\CIV} Kinematics and Phase Structure}
\label{sec:highion}

Photoionization modeling of {\MgII} absorbers has shown that {\CIV} and {\MgII}
absorption does not usually arise in the same phase (gas with the same density
and temperature).  The {\CIV} arises in a lower-density phase that produces
generally broader lines that are often aligned in velocity with components
arising in higher-density gas that gives rise to low-ionization absorption
\citep{dcb03,dcc03,mcd05}.  {\CIV} components without associated {\MgII}
are also often detected at other velocities.
There also exists a subset of strong {\MgII} systems, so-called {\CIV}-deficient
absorption, with only weak or even non-detected {\CIV} \citep{cmc00}.

The very strong quasar {\MgII} absorbers in our sample, with {\CIV} covered, share
those general features.  \cite{cmc00}
found two different categories of very strong {\MgII} absorbers: one in which
the {\CIV} absorption was also very strong (called double), interpreted as consistent
with galaxy pairs and one in which the {\CIV} was typical of that found in
$W_r(2796) < 1$~{\AA} strong {\MgII} absorbers (called DLA/{\HI}-rich), consistent
with a particularly dense region producing {\MgII} within a normal galaxy and
its halo.  Unfortunately, in only three GRB absorbers from our sample does
the {\CIV} coverage provide information.  In the $z=1.7976$ absorber toward
GRB 060607 the {\CIV} traces the {\MgII} in velocity, but two phases are
required to explain the relative strengths.  There is also a separate {\CIV}
component in this absorber without associated detected {\MgII}.  The $z=1.6015$ system
toward GRB 021004 has quite a noisy spectrum, but it appears that {\CIV}
is only detected in the bluer components of the system.  The same phenomenon is observed for the {\AlIII} associated with that system.
Finally, with the $z=1.38$ system toward GRB 021004, again the spectrum
is noisy, but it appears that the {\CIV} roughly traces the
{\MgII}.
We conclude that the GRB {\CIV} absorber profiles appear to have similar kinematic properties to
the quasar absorber {\CIV} profiles.
It is interesting to note that the \emph{dN/dz} for such highly ionized transition does not 
show significant difference between GRB and Quasars samples, as shown by \cite{tlp07}.

\subsection{Fine Structure and Metastable Transitions}
\label{sec:fine}

It is common to detect fine structure transitions in the highest-redshift -- presumed host galaxy --  absorbers of GRBs.  These
host galaxy absorbers are often DLAs, and are characterized by strong
low-ionization absorption with accompanying high-ionization absorption.
Among the six GRBs that we have studied here, GRB\,050730 \citep{sve05}, GRB\,060418 \citep{vls07} and  
GRB\,050820 \citep{pro06, bpf05, vls07}  have this 
type of absorber at the GRB redshift.  On the other
hand, for GRB\,060607, \cite{pdr08} have reported host galaxy absorbers
with only high ionization absorption features.  It is possible
the host galaxy absorption for GRB\,060607 might actually be at a lower redshift, $z=1.799$
and the detected {\CIV} absorption at $z=3.08$ might be due to material moving with positive
relative velocity along the line of sight towards the GRB at the host redshift.

In the case of our six GRBs, fine structure transitions have been reported
  in GRB\,060418, GRB\,021004, and GRB\,050730.
Ordinary intervening DLAs often have detected {\CII}$^*$, but do not have
the {\SiII}$^*$ and {\FeII}$^*$ transitions that are evident in some of these GRB
hosts.  Two major mechanisms have been discussed for producing these
fine structure lines: UV pumping or collisional excitation.  The UV
pumping mechanism, generally thought to be more likely \citep{pcb+06},
requires the host absorber to be located within the same galaxy as the
GRB..

If the excess of strong {\MgII} absorbers in GRBs are proximate to the GRB 
radiation field, we might
expect to observe fine structure transitions from them as well.  
We find that these fine structure transitions are not detected in any of
the cases here.  Fig.~\ref{fig:fig9} shows an example of the regions where we
would expect to detect {\FeII}* and {\NiII}* for the $z=1.106$ system toward GRB\,060418,
and they are not detected.  We conclude that if the excess absorption arises
in material intrinsic to the GRB, it must have properties distinct from the
``host galaxy absorbers'', for example, it might be at a closer location
but have a much higher density, as we discuss further in \S~\ref{sec:discussion}.

Absorption in the meta-stable levels of {\FeII}$^*$ and {\NiII}$^*$ was detected in the
host galaxy absorption associated with GRB\,060418, providing further evidence
that a UV pumping mechanism is favored over collisional excitation \citep{vls07}.
A distance of at least 1.7~kpc was derived for the absorber in this case, using
the variability of the fine structure and metastable transitions.
We have  examined the metastable transitions for all of the very strong {\MgII}
absorbers in our GRB sample and do not detect any of these
transitions.

\section{Discussion of Intrinsic Origin of Excess GRB Very Strong {\MgII} Absorbers}
\label{sec:discussion}

Our conclusion is that there is no significant evidence in support of
a difference between the populations of very strong {\MgII} absorbers found
toward GRBs as compared to quasars.  Any such difference
could have been used to argue for a difference  in the nature of
the GRB absorbers.  Given the outstanding issues with explanations involving
dust bias, gravitational lensing, and partial beam covering, as discussed in
\S~\ref{sec:intro}, we would expect such a difference to arise from an
origin close to the GRB.  The question now is
how we should interpret the negative result, that is, an absence of  any 
difference in the observed properties of GRB and quasar absorbers.

The most straightfoward explanation would be to say that material
ejected from the GRB environment either does not exist or does not
produce very strong {\MgII} absorption.  Indeed, the required velocities
up to $0.3$ or $0.4c$ are a challenge to models of the regions surrounding
GRBs.

We could also say that, in view of the small sample size of GRB absorbers
and the great diversity in the population of very strong {\MgII} absorbers
toward quasars, a difference would be hard to recognize.
Another option is to embrace the
similarity in physical conditions of the two population of {\MgII}
absorbers, quasars and GRB, and consider whether there could be
similarities between their environments even in a model where
GRB absorbers have an intrinsic origin.

As described in \S~\ref{sec:intro}, very strong {\MgII} absorbers toward quasars
are thought to arise from a mix of processes, with significant contributions from
dense star-forming regions and from condensations in superwinds
\citep{rtn06,ntr05,bmp06}.  Some GRBs also are thought to arise
within star forming regions.  It is intriguing to note that in the nearby
universe, the Carina nebula, an indication of the structure of the
interstellar medium near O stars, has absorption profiles very similar
to those of some of the very strong {\MgII} absorbers we are studying
\citep{wfc02}.  
Perhaps similarities in the quasar intervening and GRB absorbers
are the result of similar processes being involved.

On the other hand, it is worth noting that \citet{hfs06} have recently
identified observational evidence for GRBs being associated with
``runaway'' type massive stars, rather than those that remain near
their stellar nursery until death.  At the same time, the host
galaxies of these GRBs are Wolf-Rayet galaxies with luminous super
star clusters and galaxy-scale outflows.

Of course, perhaps the most important difference between the
intervening quasar absorbers and the immediate vicinity of the GRB is
supplied by the intense radiation of the GRB itself.  We have noted
that there is an expectation that the GRB radiation field should
destroy {\MgI} within tens of parsecs.  \cite{flv08} proposed that
high velocity components ($500$--$5000$~{\kms} relative to the GRB)
detected in {\CIV} absorption could be the result of Wolf-Rayet winds
on such small scales.  They suggest that only the cases without
detected low-ionization absorption would be consistent with an origin
in the immediate GRB environment.  The GRB radiation field should also
lead to populating the fine structure levels of various transitions at
distances up to the kiloparsec scale, thus producing absorbers
consistent with the claimed host galaxy absorption of many GRBs.
These host galaxy absorbers are close to the redshift of the GRB and
distinct from the absorbers that would produce the excess of very
strong {\MgII} absorption toward GRBs.  There may seem to be little
parameter space left for absorbers near to the GRB to have the
physical conditions required.

However, one important analogy, already mentioned in \S~\ref{sec:introintrin}, makes us pause
before dismissing an origin for the excess GRB absorbers close to the GRB.  There is a
population of intrinsic narrow absorption line systems observed in quasar spectra which
appear to have their origin in the accretion disk wind of the quasar.  In many cases these absorbers
have relativistic ejection velocities, and they have ionization parameters and kinematic
structures that are virtually indistinguishable from intervening absorbers, particularly
{\CIV} absorbers.  In some cases, related low-ionization gas is detected.  The obvious
similarity in properties between these intrinsic quasar absorbers and some intervening quasar
absorption line systems is despite the location of the intrinsic absorbers in the immediate
vicinity of the quasar radiation flux.  In order that they have the same ionization
level they have to have gas densities many orders of magnitude higher than those in
intervening absorbers.  Despite their different acceleration mechanism and environment,
their kinematic properties are somehow similar to those of the intervening absorbers.  The only way to distinguish intrinsic from intervening narrow absorption line systems is through partial
covering analysis, which is observed in only a subset of the intrinsic population \citep{mec07}.
The other members of that intrinsic population remain hidden in intervening
quasar absorption line samples.

We conclude by considering
the requirements in order that excess very strong {\MgII} absorbers seen
toward GRBs arise in the GRB environment/host galaxy.  The general similarity
between the kinematics of the {\MgII} profiles of GRB and quasar absorbers
indicates that there are similar physical processes involved which are
likely to be related to star formation.  There is a requirement that the
kinematic spread is large and, specifically, that the apparent ejection velocity is
relativistic.  Superwinds are not adequate for the latter, but 
interaction of the GRB shock wave with pre-existing 
material around it due to ``hypernova'' explosion could 
produce material with the required apparent ejection velocity.

To achieve relativistic velocities, it seems the absorbing material must
be relatively close to the GRB radiation field, yet it must still be
possible for {\MgI} to exist.  For this, it seems that high densities
and shielding would be necessary, in analogy with the situation for
intrinsic narrow absorption lines in quasars.  Also, we need to understand why
absorption is not observed in fine structure transitions in view of
the proximity to the GRB.  Clearly, these requirements are restrictive
and lead to some skepticism about an intrinsic origin.  On the other
hand, there is currently no more obvious explanation for the puzzling
excess of very strong {\MgII} absorbers toward GRBs.

In the small sample of GRB absorbers that we examined, we saw no clear
difference with the quasar absorber population, but given the wide range
of quasar absorber properties, a larger sample of GRB absorbers might 
be needed to see any such trend, if it is present.  For example, perhaps some of the GRB absorbers that we have examined have particularly small
{\MgI} equivalent widths.  Given the small number of high-quality datasets in the present sample, detailed studies of the high resolution profiles
of GRB absorbers are quite worthwhile to pursue for future GRBs. As here, future studies should take 
into account all of the observed chemical transitions and their kinematics.
Through such an expanded sample, it may be finally possible to tell if
the excess GRB {\MgII} absorbers are telling us something about intervening
absorbers or if they are telling us about some component of the
environment of the GRB.

\section{Summary of Results}
\label{sec:summary}

We compared the properties of 8 very strong {\MgII} absorbers ($W_r(2796)>1$~{\AA})
observed along 6 GRB line of sights   to the properties of absorbers at similar redshifts along 81 quasar line of sights.  Our aim was to investigate if the
reported excess of $[dN/dz]_{\rm{GRB}} \approx 3 \times [dN/dz]_{\rm{QSO}}$ found by
\cite{ppc06} is caused by excess material in the immediate proximity of the
GRB.  More specifically, \cite{ppc06} found 14 absorbers toward 14 GRBs,
where 3.8 would be expected from quasar statistics.  Similarly, we would expect
$\sim 2$ very strong {\MgII} absorbers toward the GRBs in our sample and we
find $8$.  This is not an independent datapoint from \cite{ppc06}
since there is   overlap between our samples, moreover the probability of
us finding $3$ or $4$ very strong {\MgII} absorbers in our GRB sample is
not negligible.  So, among the $8$ absorbers in our GRB sample, we might
expect about half should have an origin in the GRB environment if an intrinsic
hypothesis is valid. From Table.~\ref{table:tab2}, we see that such an interpretation of the excess
would require relativistic velocities,  to at least $v \approx 0.3c$.

Our challenge, then, was  to find differences between
an unknown subset of $4$ or $5$ of the $8$ GRB absorbers and the $27$ quasar
absorbers which themselves constitute a varied sample of intervening
environments and processes.  Nonetheless, in the intense radiation field of
a GRB, and in regions that are directly associated with active star formation, we might
expect some differences.  If the material is  directly
related to the GRB, and ejected relativistically from its immediate environment,
we might expect strong kinematic and ionization differences compared to ordinary
intervening absorbers.

In our analysis of $8$ GRB and $27$ quasar {\MgII} absorption systems we have
found no statistically significant differences:

\begin{enumerate}
\item{The kinematics of the {\MgII} $\lambda$2796 profiles do not differ between
the two samples.  This is apparent by visual comparison in Figure~\ref{fig:fig4},
and confirmed  by comparison of the kinematic spread, full velocity range, and $D$-index
quantitative measures of  kinematics, as shown in Figure~\ref{fig:fig5};}

\item{The equivalent widths of low-ionization transitions ({\MgI} and {\FeII}),
dust tracers ({\CrII} and {\MnII}), and intermediate/high-ionization transitions
({\AlIII} and {\CIV}) cover the same range of values for GRB absorbers as for
quasar absorbers.  The equivalent width distributions of {\MgII} $\lambda$2796
were not different between the two samples (among the $W_r(2796)>1$~{\AA} absorbers),
so it is meaningful to compare the absolute values of the equivalent widths of
other transitions, rather than ratios.
Many of the other transitions were only covered for the higher-redshift GRBs in our sample, and so the statistics for this comparison were
more limited.  We do note that several GRBs have particularly weak {\MgI}
compared to quasar absorbers.  If this is found to be common in a larger
sample it would be of great interest, since it is expected that {\MgI}
should not survive within 50~pc of a GRB \citep{pcb+06};}

\item{The distributions of the ratios of {\FeII} to {\MgII} equivalent widths
did not differ significantly between the two samples.  This statement applies to  saturated
{\FeII} transitions as well as to {\FeII} $\lambda$2374, which was rarely saturated
for these systems.  There is a possible evolutionary trend found in this ratio for
quasar very strong {\MgII} absorber, with few large ratios of {\FeII} to
{\MgII} at $z<1.2$;  the GRB sample is not large enough to
determine wether a similar trend holds;}

\item{We used the ratio of the equivalent widths of various {\FeII} transitions
to that of {\ZnII} $\lambda$2026 as an indication of the Fe/Zn
ratio, which indicates dust depletion since zinc depletes more readily than
iron.  There is no statistical difference in the distributions, although one
GRB absorber according to this metric ($z=1.106$ toward GRB\,060418) is confirmed to be particularly dusty
\citep{evl06};}

\item{The {\CIV} doublet is covered for three of the GRB absorbers in our sample,
and that region of the spectrum is noisy for two of the three.  In these cases we find that kinematics and strength of the {\CIV} absorption is not exceptional, compared to the range of these properties among the quasar absorbers;}

\item{Fine structure and metastable transitions were not detected for any
of the GRB absorbers in our sample.  The fine structure lines
are typically detected in association with  the absorption systems that are believed to represent 
the GRB host galaxies. Such host galaxy absorption systems are observed and interpreted as occurring in regions tens of parsecs to several kiloparsecs
from the GRB, where gas has been  excited by UV pumping from the burst prompt emission or the early
afterglow emission \citep{pcb+06, twl07,
vls07}.  The absence of
such fine structure lines in our sample suggests that a different region, or
gas with a different density, must be responsible for the intervening very strong {\MgII}
absorption seen in many GRB afterglow spectra.}
\end{enumerate}

\acknowledgments{
This work was supported by the Swift Guest Investigator Program under NASA Grant NNX08AN86G.  Work of authors Charlton and Narayanan has been funded by NASA grant NAG5-6399 and National Science Foundation grant AST-04-07138.  The work was enabled by observations made with the ESO Telescopes at Paranal Observatories, and relied heavily on the public VLT data archive provided by ESO, and on the MIDAS UVES pipeline software.}


\begin{deluxetable}{lcccc} 
\tablecolumns{5} 
\tablewidth{0pc}
\tablecaption{Observation log} 
\tablehead{ 
\colhead{GRB} & \colhead{UT start}   & \colhead{Exposure (min)} & \colhead{coverage(nm)}   & \colhead{References}}
\startdata 
021004& 2002, April 5.22 & 30 &376-498;670-852; 866-1043 & (1)(2)\\
021004& 2002, April 5.23 & 30 &302-392;473-580;576-680 & (1)(2)\\
021004& 2002, April 5.23 & 10 &565-660 & (1)(2)\\
021004& 2002, April 5.24 & 60 &452-560;568-665  & (1)(2)\\
021004& 2002, April 5.25 & 60 & 302-392;473-580;576-680 & (1)(2)\\
021004& 2002, April 5.28 & 60 &376-498;670-852;866-1043  &(1)(2)\\
\cline{1-5} 
050730& 2005, July 31.01 & 3 &302-392;473-580;576-680  &(2)(3)\\
050730& 2005, July 31.04 & 5 &373-505;665-854;864-1008  &(2)(3)\\
\cline{1-5} 
050820& 2005, August 20.29 & 30 &328-452;452-560;568-665  & (2)(4)\\
050820& 2005, August 20.32 & 30 &328-452;452-560;568-665  & (2)(4)\\
050820& 2005, August 20.35 & 40 &373-505;665-854;864-1008 &(2)(4) \\
\cline{1-5} 
050922C& 2005, September 22.98 & 50 &302-392;473-580;576-680   & (2)(4)(5) \\
050922C& 2005, September 23.02 & 50 &373-505;665-854;864-1008  & (2)(4)(5)\\
\cline{1-5} 
060418& 2006, April 14.13 & 3 &328-452;452-560;568-665  & (6)\\
060418& 2006, April 14.14 & 5 &328-452;452-560;568-665  & (6)\\
060418& 2006, April 14.15 & 10 &328-452;452-560;568-665  & (6)\\
060418& 2006, April 14.15 & 20 &328-452;452-560;568-665 & (6)\\
060418& 2006, April 14.17 & 40 &328-452;452-560;568-665  & (6)\\
060418& 2006, April 14.20 & 80 &376-498;670-852;866-1043& (6) \\
\cline{1-5} 
060607& 2006, June 7.22 & 3 &328-452;452-560;568-665 & (2)(4)\\
060607& 2002, June 7.23 & 5 &328-452;452-560;568-665 &(2)(4) \\
060607& 2002, June 7.23 & 10 &328-452;452-560;568-665  &(2)(4)\\
060607& 2002, June 7.24 & 20 &328-452;452-560;568-665  &(2)(4)\\
060607& 2002, June 7.25 & 40 & 328-452;452-560;568-6650 &(2)(4)\\
060607& 2002, June 7.28 & 80 & 376-498;670-852;866-1043 &(2)(4)\\
060607& 2002, June 7.28 & 42 &376-498;670-852;866-1043  &(2)(4)\\

\enddata 
\tablecomments{References are numbered as follows: (1) \cite{fdl05}, (2) \cite{flv08}; (3) \cite{dfm07}; (4) \cite{ssv07};  (5) \cite{pwf08}; (6) \cite{vls07}}
\end{deluxetable}

\begin{deluxetable}{lccc} 
\tablecolumns{4} 
\tablewidth{0pc}
\tablecaption{Systems} 
\tablehead{ 
\colhead{Object} & \colhead{$z_{\rm{obj}}$ }   & \colhead{$z_{\rm{sys}}$ } & \colhead{$\Delta v/c$    }}

\startdata 
GRB060418 (1)&	1.49&	0.6021&	0.41\\
GRB060418 (2)&	1.49&	0.6554&	0.38\\
GRB060418 (3)&	1.49&	1.1066&	0.16\\
GRB021004 (1)&   2.328&  1.3800 &   0.32 \\
GRB021004 (2)&	2.328&	1.6015&	0.24\\
GRB050820 (1)&	2.612&	0.6910&	0.64\\
GRB050820 (2)&	2.612&	1.4280&	0.37\\
GRB060607&	3.08&	1.7996&	0.36\\
\cline{1-4}
3C336 (1)			& 0.9273	&  0.6561	&0.1505\\
3C336 (2)			& 0.9273	& 0.8913	&0.0188\\
Q1229-021		& 1.038	& 0.3951 & 0.3618 \\
Q1127-145 		& 1.18	& 0.3127   & 0.4677\\
Q1629+120		&1.795	& 0.9002	&0.3677\\
Q0328-272		&1.816	&1.1228		&0.2753\\
Q1331+170		&2.084	&1.7766		&0.1046\\
HE1341-1020		&2.134	&1.2767	&0.3091\\
PKS0237-23 (1)		&2.224	&1.3651		&0.3003\\
PKS0237-23 (2)		&2.224	&1.6723		&0.1855\\
Q0551-3637		&2.318	&1.9609		&0.1134\\
Q0109-3518		&2.35	&1.3496		&0.3405\\
HE1122-1648		& 2.4	& 0.6822	&0.6066\\
HE2217			&2.406	&1.6922		&0.2309\\
Q0453-423 (1)		&2.66	& 0.7261 	&0.6361\\
Q0453-423 (2)		&2.66	& 1.1498	&0.3642\\
Q0100+130 (1)		&  2.681	& 0.2779 & 0.7848\\
Q0100+130 (2)		&  2.681	&  1.7969 & 0.2679\\
Q0002 (1)			&2.76	&0.8366		&0.6147\\
Q0002 (2)			&2.76	&2.3019		&0.1292\\
Q1151+068		&2.762	&1.8191		&0.2808\\
Q0112+300		&2.81	& 1.2452	& 0.4800\\
Q0130-4021		&3.03	&0.9315 	&0.6264\\
HE0940			&3.083	&1.7891		&0.3636\\
CTQ0298			&3.37	&1.0387 	&0.6427\\
Q1418-064		&3.689	&1.4578		&0.5689\\
Q1621-0042		&3.7		&1.1335		&0.6583\\
\enddata 
\label{table:tab2}
\tablecomments{\footnotesize {List of GRB and QSO systems  identified, in increasing object redshift order, with rest frame EW($\lambda2796>1$  \AA). The redshift of the object ($z_{\rm obj}$) is the quasar or GRB redshift, estimated by emission lines (in the QSO case) or multiple absorption features associated with the host galaxy redshift (for the GRB). The system redshifts are estimated combining information from different features aligned with the strong \MgII systems. The last column reports the apparent ``ejection'' velocity, estimated assuming that the absorbing system is actually moving at a positive velocity respect the quasar or the GRB host galaxy.} }  
\end{deluxetable}



\begin{deluxetable}{lccccccccccccc} 

\tabletypesize{\scriptsize}
\rotate

\tablecolumns{12} 
\tablewidth{-9in}
\tablecaption{Equivalent widths of the transitions} 
\tablehead{ 
\colhead{System} & \colhead{$z_{\rm{sys}}$ }   & \colhead{MgII2796 } & \colhead{MgII2803} & \colhead{MgII2853} & \colhead{FeII2383} & \colhead{FeII2374 }   & \colhead{FeII2587 } & \colhead{FeII2600} & \colhead{MnII2577} &  \colhead{MnII2594} & \colhead{MnII2606 } }
\startdata 
GRB060418 (1)	& 0.6021 &	$1.366(26)$ & $1.242(24)$ &...&$0.932(25)$& ... & $0.648(12)$ & $0.831(09)$ & $0.159 (14) $ & $0.099(114) $ & $ 0.055(07)$\\
GRB060418 (2)	& 0.6554 &	$ 1.052(18)$ & $0.836(21)$ &$0.053 (10)$& $0.687(26)$&$0.117(13)$& $0.295(17)$ & $0.436 (19)$ & $<0.014$ & $<0.027 $ & $<0.023 $ \\
GRB050820 (1)	& 0.6910 &	$2.950(03)$ & $2.350 (05)$ &$1.435(45)$& $0.991(23)$& $0.449(18)$ & $0.743 (37)$ & ... & $1.329 (23) $ & $<0.065 $ & $< 0.085 $\\ 
GRB060418 (3)	& 1.1066 &	$ 1.855(13)$ & $1.447(14)$ &$0.503 (12)$&$...$&$1.115(14)$& $0.486(15)$ & $0.699 (10)$ & $1.112 (12) $ & $0.106 (10) $ & $ 0.090 (09)$ \\ 
GRB021004 (1)	& 1.3800 &	$1.759(23)$ & $1.475(23)$ &$0.120(15)$& $0.859(22)$&$0.559(18)$ & $0.617(19)$ & $0.806(20)$ & $<0.023 $ & $0.294(24) $ & $ <0.090$ \\
GRB050820 (2)	& 1.4280 &	$1.170(03)$ & $1.070(03)$ &$0.231(01)$& $0.485(10)$&$0.208(08)$ & $0.365(08)$ & $0.501(09)$ & $<0.008 $ & $<0.017 $ & $ <0.006$ \\
GRB021004 (2)	& 1.6015 &	$1.507(47)$ & $1.207(26)$ &$<0.159 $& $0.786(23)$&$0.071(13)$ & $0.306(12)$ & $0.677(17)$ & $<0.029 $ & $<0.018$ & $ <0.018$ \\
GRB060607 	& 1.7996 &	$1.898(10)$ & $1.603(14)$ &$<0.011$& ...& ... & $0.109(06)$ & $0.859(09)$ & $<0.005 $ & $<0.011 $ & $ 0.016(10)$ \\

\cline{1-12}
Q0100+130 (1)	   & 0.27795 &    $2.316 (13)$  & $ 2.246 (17)$ &$0.988   (11)$&...&... & ... & ... & ... & ... &... \\
Q1127-145		& 0.3127  &   $1.706 (04)$ & $1.651 (05)$ & $1.027 (05) $ &$ 1.105(16)$ & $0.843 (19) $& $1.191(05)$ & $1.321(06)$ & $0.264 (07)$ & $0.199 (08)$ & $0.159 (09)$ \\ 
Q1229-021	   & 0.3951   &     $2.064(06)$  & $ 1.792 (05)$ &$0.619 (05)$& ...& ... & $0.867(08)$  & $1.242 (09)$&$0.282(09)$ & $0.224(08)$ &$0.192 (08)$ \\
3C336  (1) & 0.6561 &    $1.428 (09)$  & $ 1.290 (09)$ &$0.213(09)$ & $1.055 (09)$&$0.480 (10)$ &$0.797(10)$ & $1.112(10)$ & $0.061(09)$ &$0.037(08)$ &$<0.009$ \\
HE1122	   & 0.6822 &    $1.712(02)$  & $1.569 (02)$ &$1.431 (03)$&...&$0.397(01)$ &$0.816 (01)$ & $1.277(01)$ & $0.052(01)$& $0.029(01)$ &$0.020(01)$ \\
Q0453 (1)  & 0.7261 &    $1.366(02)$  & $ 1.268 (02)$ &$0.455 (02)$&...&... & $0.710(01)$ &$1.052(01)$ & $0.309(02)$&$0.078(02)$ &$0.055(02)$ \\
Q0002 (1)	   & 0.8366 &    $4.431(02)$  & $ 3.954 (02)$ &$1.586(02)$&$3.093(03)$&... & $1.833(03)$ &$ 2.994(02)$ & $0.052(02)$ & $0.028(01)$ &$0.021(01)$ \\
3C336 (2)	   & 0.8913 &    $1.519(06)$  & $ 1.392 (06)$ &... &...&$0.397(09)$ & $0.744 (07)$ & $1.114(07)$& $0.031 (05)$ & $0.026 (04) $ &$<0.003$ \\
Q1629+120	   & 0.90025 &    $1.022 (07)$  & $ 0.772 (05)$ &$0.184 (06)$& ...&... &$0.383(07)$  & $0.702(09) $ & $0.014(03)$ & $0.018(04)$ &$<0.003$ \\
Q0130-4021	   & 0.9315   &    $1.230(06)$  & $ 0.890(01)$ &...&...&... & $0.152(05)$ & $0.289(04)$ & $<0.002$ & ... &$<0.002$ \\
CTQ0298	       & 1.0387  &    $1.555 (06)$  & $ 1.279(06)$ &$0.118 (05)$&$1.685(09)$&... & $1.213(08)$ &$0.744(04)$ & ... & ... &$<0.004$ \\
Q0328-272	   & 1.1228 &    $1.219(11)$  & $ 1.186 (10)$ &$0.081(06)$&$0.912(08)$&$0.394(08)$ & $0.599(07)$ & $0.906(07)$ & $<0.007$ &$<0.005$ &$<0.004$ \\
Q1621-0042	   & 1.1335 &    $3.168 (04)$  & $ 2.711 (04)$ &$0.397(04)$&...&... & ... & ... & $<0.002$ & ... &... \\
Q0453 (2)	   & 1.1498 &    $4.445(02)$  & $ 3.987 (02)$ &$1.521(02)$&$3.573(02)$&$1.778(02)$ & $2.685(02)$ & $3.653(01)$ & $0.412(02)$ & $0.309(01)$ &$0.222(02)$ \\
Q0112+0300	   & 1.2452 &     $3.099(19)$  & $ 2.755(14)$ &$0.953 (22)$&$2.187(15)$ &$0.960(18)$ & ... & $0.886(44)$ & ... & ... &$<0.012$ \\
HE1341	  		 & 1.2767 &    $1.450 (01)$  & $ 1.330 (02)$ &$0.300(01)$& $0.802(07)$&$0.300(01)$ & $0.540(01)$ & $0.814(07)$ & $<0.002$ &$<0.003$ &$<0.002$ \\
Q0109	   	& 1.3496 &    $1.980 (02)$  & $ 1.582(02)$ &$0.286(02)$&$0.839(02)$&$0.124(02)$ & $0.352(02)$ & $0.798(02)$ & $<0.001$ & $<0.002$ &$<0.002$ \\
PKS0237	(1)   & 1.3651 &    $1.856 (01)$  & $ 1.646(01)$ &$0.268(02)$&$0.885(01)$&$0.206(01)$ & $0.409(01)$ & $0.870(01)$& $0.008(01)$ & $0.007(01)$&$0.003(01)$ \\
Q1418-064	   & 1.458 &    $2.155 (09)$  & $ 1.843 (11)$ &$0.159 (07)$&...&... & ... & ... & ... & ... &... \\
PKS0237	(2)   & 1.672334 &    $1.283 (01)$  & $ 1.071(02)$ &...&$0.519(01)$&$0.256(01)$ & $0.425(01)$ & $0.529(01)$& $0.030(01)$ & $0.031(01)$ &$0.025(01)$ \\
HE2217	   & 1.6922 &    $1.681 (01)$  & $ 1.235 (01)$ &$0.051(01)$&$0.332(01)$&$0.045(01)$ & $0.085(01)$ &$0.361(01)$ & $<0.004$ & $<0.002$ &$<0.002$ \\
Q1331+170	   & 1.777 &    $1.207 (02)$  & $ 1.102 (03)$ &$0.334(03)   $&...& ... & $0.701(02)$ & $0.837(02)$ & $0.054(03)$ & $0.519(04)$ &... \\
HE0940	   & 1.789 &    $1.129 (02)$  & $ 0.808 (02)$ &$0.125(01)$&$0.271(01)$&$0.036(01)$ & $0.109 (01)$& $0.259(01)$ & ... & ... &$<0.001$ \\
Q0100+130 (2)	   & 1.79698 &   $1.014(06)$ & $0.745(05)$ & $0.100(04)$ & ... &... & $0.140(05)$ &$0.197 (04)$ &$ 0.0245(01)$& $<0.010$ &  $ <0.007 $\\
Q1151+068	   & 1.8191 &    $1.069 (16)$  & $ 0.726 (14)$ &$0.100(07)$&$0.242(05)$&... & $0.169(08)$ & $0.226(11)$& $<0.006$ & ...&$<0.002$ \\
Q0551-3637	   & 1.9609 &    $4.747(09)$  & $ 4.365(01)$ &$0.979 (11)$&$3.243(07)$&$1.632(07)$ & $1.632(07)$ &$2.818(08)$ & ... & ... &... \\
Q0002 (2)	   & 2.302 &    $1.738 (02)$  & $ 1.459 (02)$ &$0.242 (02)$&$0.657(01)$&$0.187(01)$ & ... & ... & $0.026(01)$ & ... &... \\

\enddata 
\label{table:tab3}
\tablecomments{This table reports all the species searched in this study. Equivalent widths are reported at $5\sigma$ confidence level. Error values, in m\AA, are quoted in parenthesis. Upper limits are also
reported. }
\end{deluxetable}

\clearpage
\topmargin 0.0in

\begin{deluxetable}{lcccccccccccccccc} 
\tabletypesize{\scriptsize}
\tablecolumns{17} 
\tablewidth{0pt}
\rotate
\label{table:tab4}
\tablecaption{Equivalent widths of the transitions}

\tablehead{ 

	\colhead{System} & \colhead{$z_{\rm{sys}}$ }   & \colhead{CrII2056} & \colhead{CrII2062} & \colhead{CrII2066 }   & \colhead{AlIII1855 } & \colhead{AlIII1863} & \colhead{NiII1710} & \colhead{NiII1752} & \colhead{CII}   & \colhead{CIV} & \colhead{CIV}  }
\startdata 
GRB060418 (1)	& 0.6021 &	... & ... &...& ...& ... & ... & ... & ... & ... & ...  \\
GRB060418 (2)	& 0.6554 &	... & ... &...& ...& ... & ... & ... & ... & ... & ...   \\
GRB050820 (1)	& 0.6910 &	... & ... &...& ...& ... & ... & ... & ... & ... & ...  \\
GRB060418 (3)	& 1.1066 &	$<0.013$& $<0.020$&$0.122(15)$& 0.$299(17)$& $0.164(14)$ &$0.092(18)$ &$0.088(15)$ & ... & ... & ...\\
GRB021004 (1)	& 1.3800 &	$<0.0334$& $<0.019$ &...& $<0.034$& $<0.018$ & $<0.048$ &$<0.027$ & ... & ... & ...  \\
GRB050820 (2)	& 1.4280 &	... & ... &...& ...& ... & ... & ... & ... & ... & ... \\
GRB021004 (2)	& 1.6015 &	... & ... &...& ...& ... & ... & ... & ... & ... & ...  \\
GRB060607 	& 1.7996 &	... & ... &...& ...& ... & ... & ... & ... & ... & ... \\

\cline{1-12}
Q0100+130 (1)	   & 0.27795 &   ...  & ...&...&...&... & ... & ... \\
Q1127-145		& 0.3127  &   ... & ...& ... &... & ...& ...  & ... & ... & ... & ... \\
Q1229-021	   & 0.3951   &     ...  & ... & ... & ...& ... & ...  & ...&... & ...  &...   \\
3C336  (1) 		& 0.6561 	   &    ...  & ... &... & ...&...	 &...	 & ... & ... &...	&...	\\
HE1122	   & 0.6822 &    ... & ... &...& ...&... &... & ... & ...& ... &... \\
Q0453  (1) & 0.7261 &    ...  & ... &...&...&... & ...&... & ...&... &... \\
Q0002 (1)	   & 0.8366 &    ...  & ...  &... &$2.146(08)$ &$1.929(07)$ & $<0.037$ &$ <0.002$ & ...  & ...  &...  \\
3C336 (2)	   & 0.8913 &    $<0.006$  & $<0.005$ &$<0.007$&...&$0.066(08)$ & ... & ...& ... \\
Q1629+120	   & 0.90025 &    ...  &... &...& ...&$<0.006$ &...  & $<0.004 $ & ... &...&...\\
Q0130-4021	   & 0.9315   &    ...  & $ <0.004$ &$<0.003$&...&... & ...& ... & ... & ... &... \\
CTQ0298	       & 1.0387  &    $<0.011$  & ... &$<0.006$&...&... & ... &... & ... & ... &...  \\
Q0328-272	   & 1.1228 &   ...   & $ <0.002$&$<0.002$&$0.100(07)$&$0.081(07)$ & $<0.005$ & $<0.005$ &... &... &... \\
Q1621-0042	   & 1.1335 &    ...  & ...  &$<0.004$& ...&... & $<0.011$ &$<0.023$ & ...  & ... & ...  \\
Q0453 (2)	   & 1.1498 &   ...  & ... &...&...&... & $<0.002$ & ... & ...& ... &...\\
Q0112+0300	   & 1.2452 &     ...  & ... &...&... &... & ... & $<0.009$ & ... & ... &... \\
HE1341	  		 & 1.2767 &    $<0.002$  & $ <0.003$ &$<0.003$& $0.265(08)$&$0.147(08)$ & $<0.003$ & $<0.005$ & ... &... &... \\
Q0109	   	& 1.3496 &    $<0.002$  & $ <0.002$ &$<0.001$&$0.141(01)$&$0.069(02)$ & ... & ... &... &...  &$<0.609$ \\
PKS0237	(1)   & 1.3651 &    $<0.001$  & $ <0.001$ &$<0.002$&$0.396(02)$&$0.367(02)$ & $<0.002$ & ... & ...  & $1.604(03)$&$1.409(02)$  \\
Q1418-064	   & 1.458 &    ...   & ...  &... & ...&... & ... & ... & ... & ... &... \\
PKS0237	(2)   & 1.672334 &    $0.021 (01)$  & $ 0.016 (01)$ &$0.008(01)$&$0.185(01)$&$0.109(01)$ & $0.026(01)$ & $0.022(01)$& ...  & ...  &... \\
HE2217	   & 1.6922 &    $0.003(01) $  & $ <0.001$ &$<0.001$&$0.084(01)$&$0.056(01)$ & $<0.001$ &$<0.001$ & ...  & $1.067(01)$ &$0.837(01)$ \\
Q1331+170	   & 1.777 &    ...  & ...  &... &...& ... & ...  & ...  &...  &... & ...  \\
HE0940	   & 1.789 &   ...   &...  &... &$0.086(01)$&$0.041(01)$ & ... & $<0.001$ & ... & ... &...   \\
Q0100+130 (2)	   & 1.79698 &  ... & ... & ... & ... &...&...&... & $0.887(07)$&$<0.687$ &$<0.505$\\ 
Q1151+068	   & 1.8191 &   ...   & ...  &... &$0.038(03)$&$0.021(03)$ & $<0.004$ & $<0.001$& ...  & ...&...\\
Q0551-3637	   & 1.9609 &    ...  & ...  &... &$1.267(05)$&$0.799(06)$ & $0.145(05)$ &$0.113(04)$ &$2.777(08)$ & $1.825(09)$ &$1.557(11)$ \\
Q0002 (2)	   & 2.302 &    $0.006(01)$  & $ <0.001$ &$0.387(01)$&$0.388(06)$&$0.023(01)$ & ... & ... & $1.098(01)$ & $1.099(01)$ &$0.928(01)$ \\

\enddata 
\tablecomments{This table reports all the species searched in this study. Equivalent widths are reported at $5\sigma$ confidence level. Error values, in m\AA, are quoted in parenthesis. Upper limits are also
reported. }
\end{deluxetable}
\clearpage

\topmargin 0.0in

\begin{deluxetable}{lcccccccc} 
\tabletypesize{\scriptsize}
\tablecolumns{9} 
\tablewidth{0pt}
\rotate
\label{table:tab5}
\tablecaption{Equivalent widths of the transitions}

\tablehead{ 

	\colhead{System} & \colhead{$z_{\rm{sys}}$ }   & \colhead{ZnII2026} & \colhead{ZnII2063}  & \colhead{SiIV} &\colhead{SiIV} & \colhead{OI }   & \colhead{NV } & \colhead{NV} }
\startdata 
GRB060418(1)	& 0.6021 & ... &...& ... &... & ...&...& ... \\
GRB060418 (2)	& 0.6554 &  ... &...& ... &... & ...&...& ... \\
GRB050820 (1)	& 0.6910 & ... &... & ... &... & ...&...& ...\\
GRB060418 (3)	& 1.1066 &$0.369(21)$&$0.102(13)$  & ... &... & ...&...& ... \\
GRB021004 (1)	& 1.3800 &$<0.019$ &$<0.024$ & ... &... & ...&...&...\\
GRB050820 (2)	& 1.4280 &$<0.006$ &$<0.003$ & ... &... & ...&...& ... \\
GRB021004 (2)	& 1.6015 & $<0.016$&$<0.031$& ... &... & ...&...& ... \\
GRB060607 	& 1.7996 &	... &...& ... &... & ...&...& ...  \\
\cline{1-9}
Q0100+130 (1)	   & 0.27795 &... & ...& ... & ... &... &... & ...\\
Q1127-145		& 0.3127  &... &  ... & ... &  ... & ...& ...&...\\
Q1229-021	   & 0.3951   &... & ... &... & ...   & ... & ... &...\\
3C336  (1) 		& 0.6561 	   &... & ...&... & ... &  ... &...& ...\\
HE1122	   & 0.6822 &  ... & ... &... & ... &...&...&...\\
Q0453 (1)  & 0.7261 &$<0.173$& ...&...& ...&...&... &...\\
Q0002 (1)	   & 0.8366 &	... & ...&... & ...& ... & ...& ...\\
3C336 (2)	   & 0.8913 &$<0.009$ & $<0.007$&... &... &...& ...&...\\
Q1629+120	   & 0.90025 &   $0.026(04)$ & ...&...& ... &...&...&...\\
Q0130-4021	   & 0.9315   &... & $<0.004$ &... & ...& ...& ...& ...\\
CTQ0298	       & 1.0387  &... & ...&... & ... & ...&... &...\\
Q0328-272	   & 1.1228 & $0.020(03)$ & $<0.002$  &... &... & ...& ...&...\\
Q1621-0042	   & 1.1335 & ... & ... &... & ...  & ...& ...&...\\
Q0453 (2)	   & 1.1498 &  ... & ... &... & ... &... & ...& ...\\
Q0112+0300	   & 1.2452 & ... & ...&... & ... & ...&... &...\\
HE1341	  		 & 1.2767 &$<0.004$ &$<0.004$ &... &...&...&...&...\\
Q0109	   	& 1.3496 &$<0.001$ & $<0.003$&$<0.763$ & ...&...& ...& ...\\
PKS0237	(1)   & 1.3651 &$<0.002$ & $<0.002$ &...  & ... &...&... &...\\
Q1418-064	   & 1.458 &... & ...&... & ... &...&...&...\\
PKS0237 (2)	   & 1.672334 &$0.013(01)$ & $0.006(01)$  &... &... & ... & ...&...\\
HE2217	   & 1.6922 &    $<0.001$ & $<0.001$ &...  & $0.599\pm0.001$&...&...&...\\
Q1331+170	   & 1.777 & ... & ...&... & ...&$0.491\pm 0.002$ & $0.153\pm 0.006$ & $0.049\pm 0.004$ \\
HE0940	   & 1.789 &$<0.002$ & ... &...  & ...& ...  & ...  & ...\\
Q0100+130 (2)	   & 1.79698 &... & ...& $1.028\pm 0.008$ &  $0.503\pm0.005$& ... & ... &...\\ 
Q1151+068	   & 1.8191 &$<0.001$ & ...  &...  & ... & ...&... & ...\\
Q0551-3637	   & 1.9609 &$0.311(04)$ & $0.238(04)$ &$1.312\pm 0.009$ & $0.991\pm 0.010$ & $2.189\pm0.006$ & ... &...\\
Q0002 (2)	   & 2.302 & $0.005(01)$ &$<0.001$ &$0.806\pm0.001$  &$1.128\pm 0.001$ & ... & ... &...\\
\enddata 
\tablecomments{This table reports all the species searched in this study. Equivalent widths are reported at $5\sigma$ confidence level. Error values, in m\AA, are quoted in parenthesis. Upper limits are also
reported. }
\end{deluxetable}
\clearpage

\topmargin 0.0in
\oddsidemargin 0.0in



\figsetstart
\figsetnum{1}
\figsettitle{GRB sample}

\figsetgrpstart
\figsetgrpnum{1.1}
\figsetgrptitle{Absorber at  $z = 0.6021$ along GRB 060418 ($z = 1.49$) line of sight.  Transitions are aligned in velocity space, with the {\MgII} $\lambda2706$ transition centered at the median optical depth. Figures 1.1 -- 1.8 are available in the online version of the Journal.}
\figsetplot{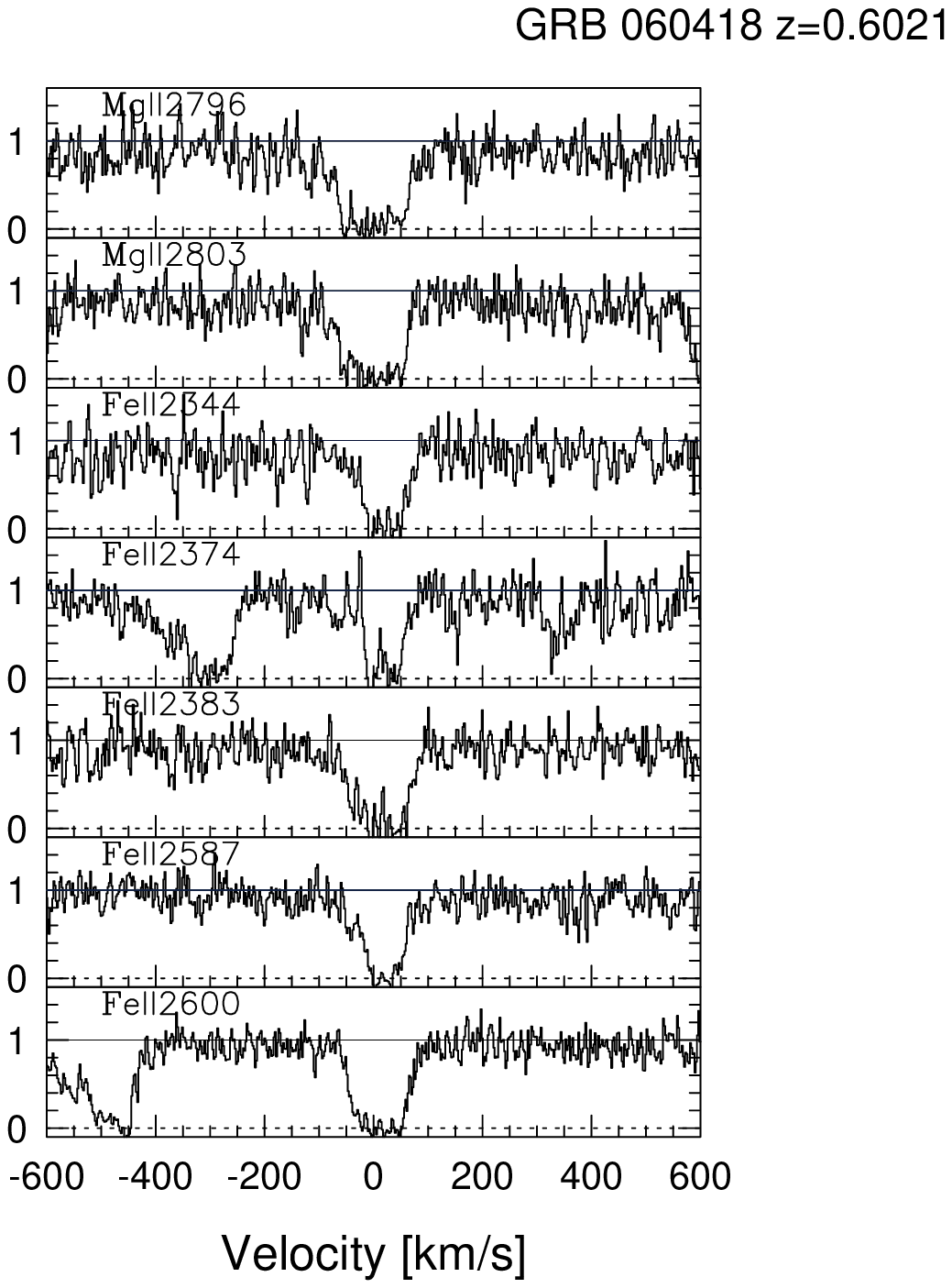}
\figsetgrpnote{GRB systems sample}
\figsetgrpend

\figsetend

\begin{figure}
\figurenum{1}
\plotone{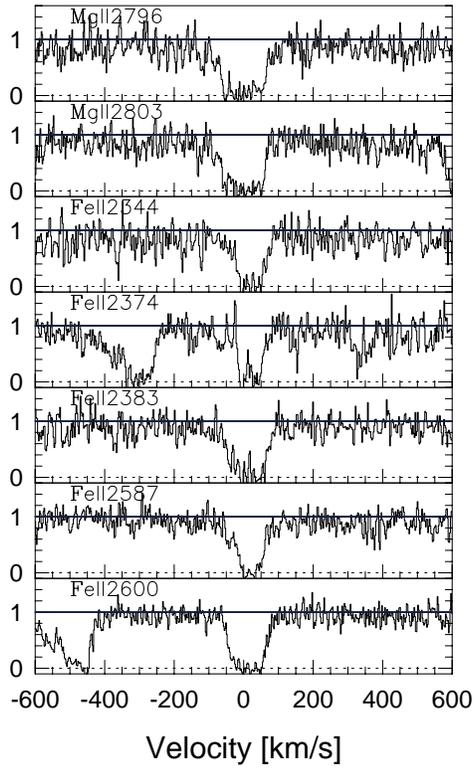}
\caption{GRB systems sample}
\end{figure}


\figsetstart
\figsetnum{2}
\figsettitle{GRB sample}

\figsetgrpstart
\figsetgrpnum{2.1}
\figsetgrptitle{Absorber at $z = 0.2779$ along QSO0100+0130 ($z = 2.681$) line of sight.Transitions are aligned in velocity space, with the {\MgII} $\lambda2706$ transition centered at the median optical depth.Figures 2.1 -- 2.27 are available in the online version of the Journal.}
\figsetplot{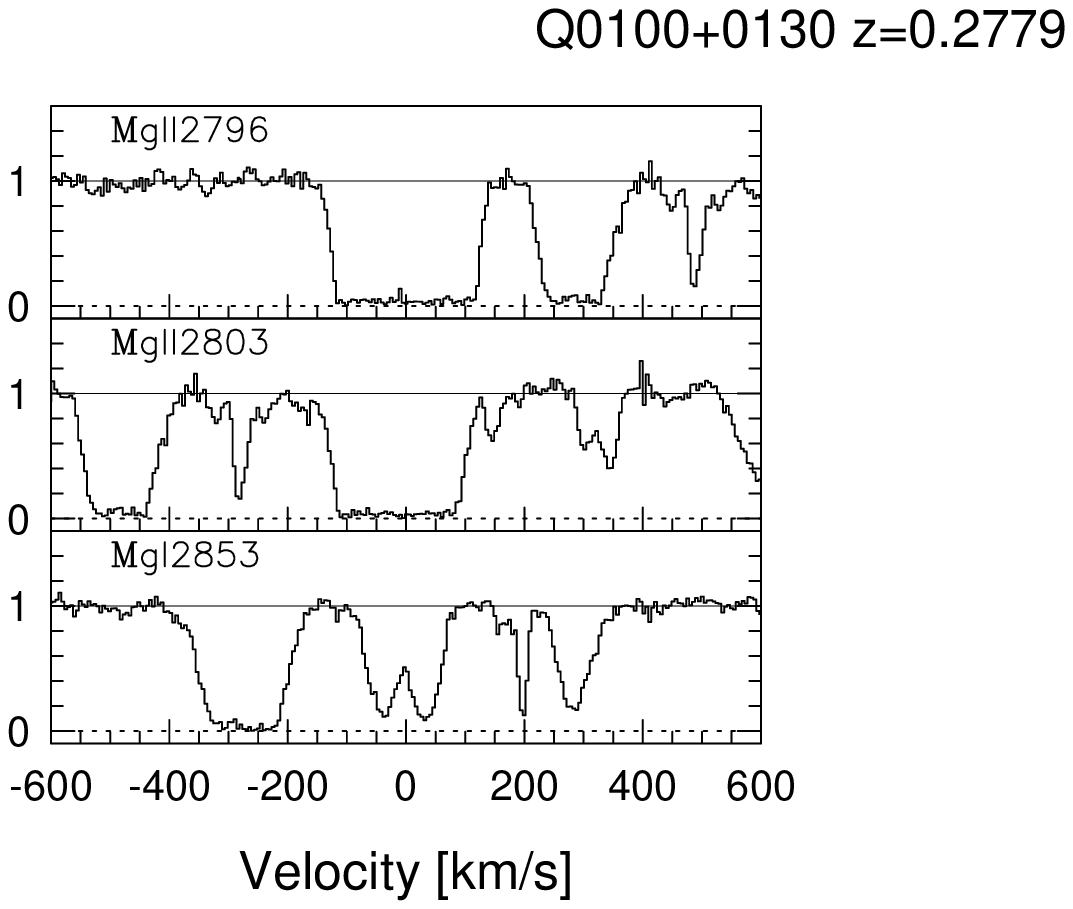}
\figsetgrpnote{QSO systems sample}
\figsetgrpend

\figsetend

\begin{figure}
\figurenum{2}
\plotone{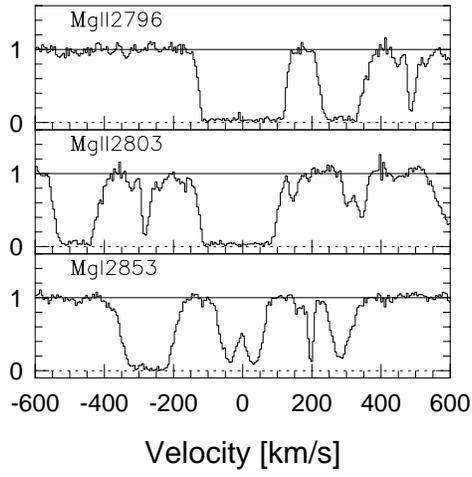}
\caption{QSO systems sample}
\end{figure}


\begin{figure*}
\figurenum{3}
 
\epsscale{0.9}
\plotone{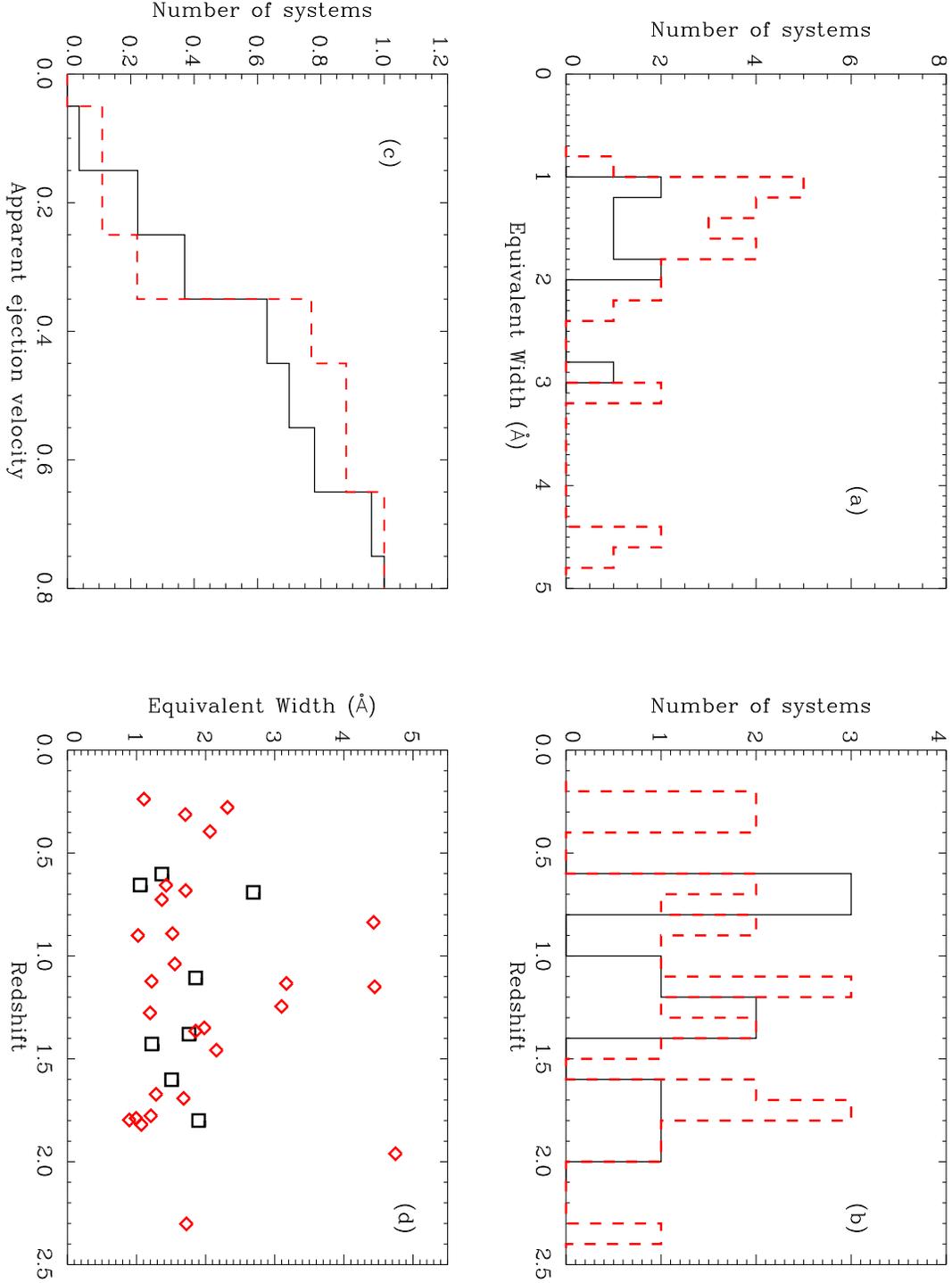}
\label{fig:fig3}
\caption{\footnotesize{Observed properties of the GRB (solid black histogram) and
  quasar (dashed red histogram) strong {\MgII} absorption systems: (a)
  Distribution in {\MgII}(2796) equivalent width; (b) Distribution in
  redshift; (c) Cumulative distribution in $\Delta v/c$, the relative or
  ``ejection'' velocity of the absorber relative to the target object normalized
  to one; (d) Two-dimensional distribution in {\MgII}(2796) rest frame equivalent width
  and redshift for GRB (black square) and quasar (red diamond)
  absorption systems.  All fundamental properties of the two samples
  exhibit a similar distribution over a similar range. }}
\end{figure*}
 \clearpage


\begin{figure*}
\figurenum{4}
\epsscale{0.9}
\plotone{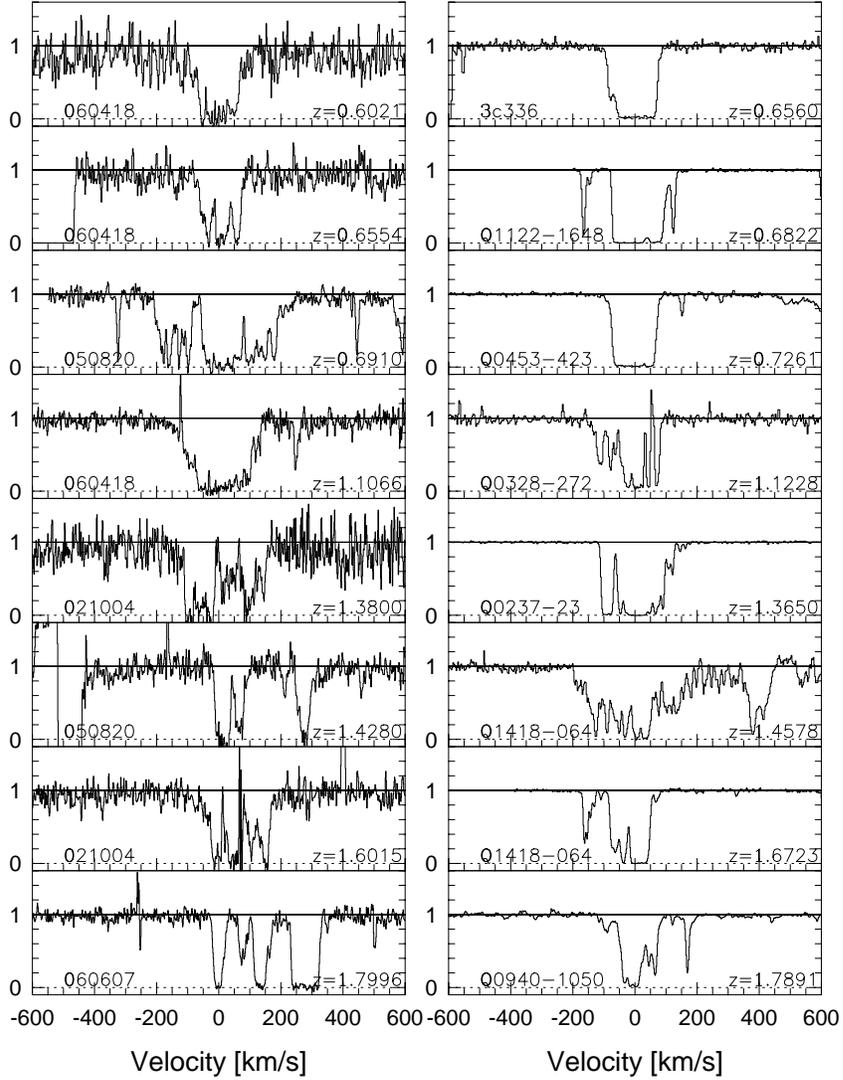}

\label{fig:fig4}
\caption{\footnotesize{{\MgII} $\lambda 2976$ line profiles for all GRB (left) and
 for eight  selected quasar (right) intervening systems. Each panel provides the
  name of the target object and the redshift of the intervening
  absorption system. No obvious systematic difference has been found
  in the absorption profiles of the two samples; in particular, both
  samples show saturated absorption features and multiple components
  with similar ranges in velocity.  Quantitative metrics derived from
  these kinematic profiles are compared in \S\ref{sec:kinematics}.
  Fig.~\ref{fig:fig5} confirms the absence of systematic differences
  between the two populations.}}
\end{figure*}

\clearpage


\begin{figure*}
\figurenum{5}
\epsscale{0.85}
\plotone{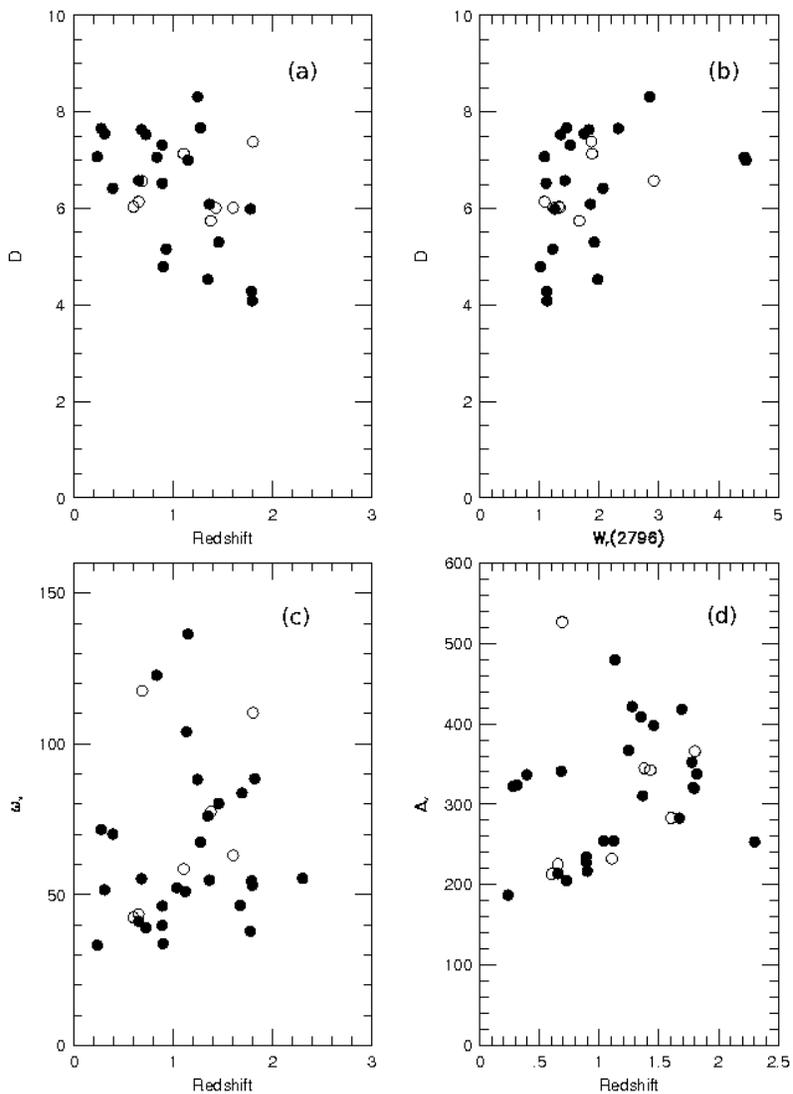}
\label{fig:fig5}
\caption{\footnotesize{Kinematic quantities describing the GRB (open symbols) and quasar
  (filled symbols) strong {\MgII} absorbers in our study:  (a)
  D-index \citep{evl06} versus redshift; (b) D-index versus
  {\MgII}(2796) rest frame equivalent width; (c) Velocity spread $\omega_v$
  versus redshift; and (d) Total velocity coverage $\Delta_v$ versus
  redshift.  There is no significant difference in either the range of
  the properties, or the distribution of properties of individual
  systems within the overall range, between the two samples, as
  confirmed in each case by a K-S test. }}
\end{figure*}
\clearpage

\begin{figure*}
\figurenum{6}

\epsscale{1.0}
\plotone{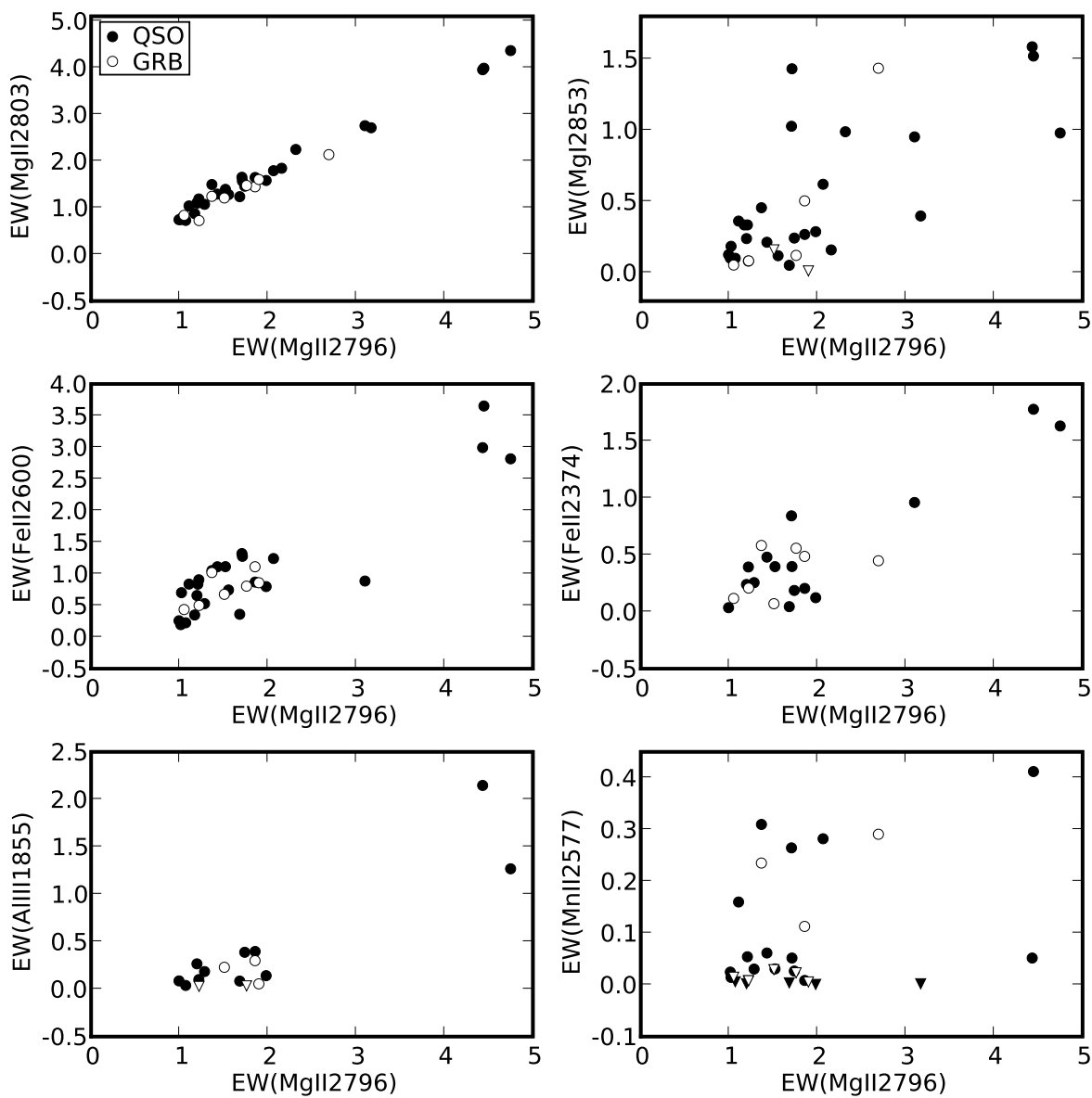}
\caption{Rest-frame equivalent widths for different species compared
  to associated {\MgII} $\lambda2706$ rest-frame equivalent width for
  GRB (open symbol) and quasar (filled symbol) absorption systems in
  our study. Triangles represent upper limits. No particular contrast
  between the GRB and quasar samples is evident. The
  detection of \MgI\ in some GRB absorbers suggests that the absorbing
  gas cannot be located within 50~pc of the GRB afterglow
  \citep{pcb+06}. }
\label{fig:fig6}
\end{figure*}
\clearpage


\begin{figure*}
\figurenum{7}
\epsscale{1.0}
\rotate

\plotone{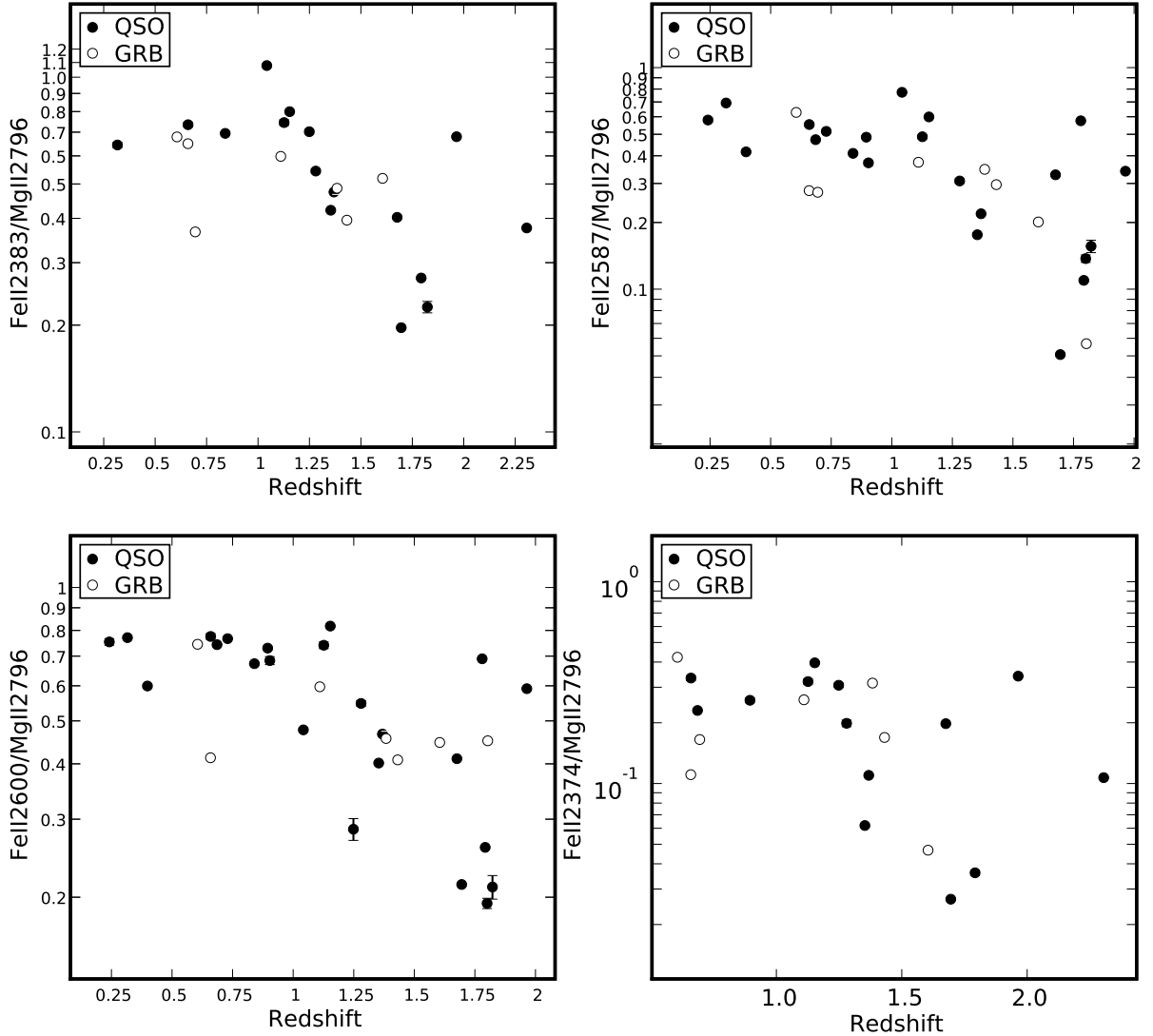}

\label{fig:fig7}
\caption{\FeII/\MgII~equivalent width ratios for different
  transitions, as observed in GRB (open symbol) and quasar (filled
  symbol) absorption systems from our study.  Errors are typically
  smaller than the symbol size. The quasar absorbers exhibit a possible evolution in
  \FeII/\MgII ~equivalent width ratio with redshift, which cannot be
  confirmed for the GRB sample because of limited statistics.  If
  present, such an evolution in \FeII/\MgII ~might suggest that
  $\alpha$-enhancement plays a dominant role at high redshift while
  type~Ia supernovae are responsible for a greater fraction of metals
  at $z< 1.2$. }
\end{figure*}
\clearpage


\begin{figure*}
\figurenum{8}
\epsscale{1.0}
\rotate
\plotone{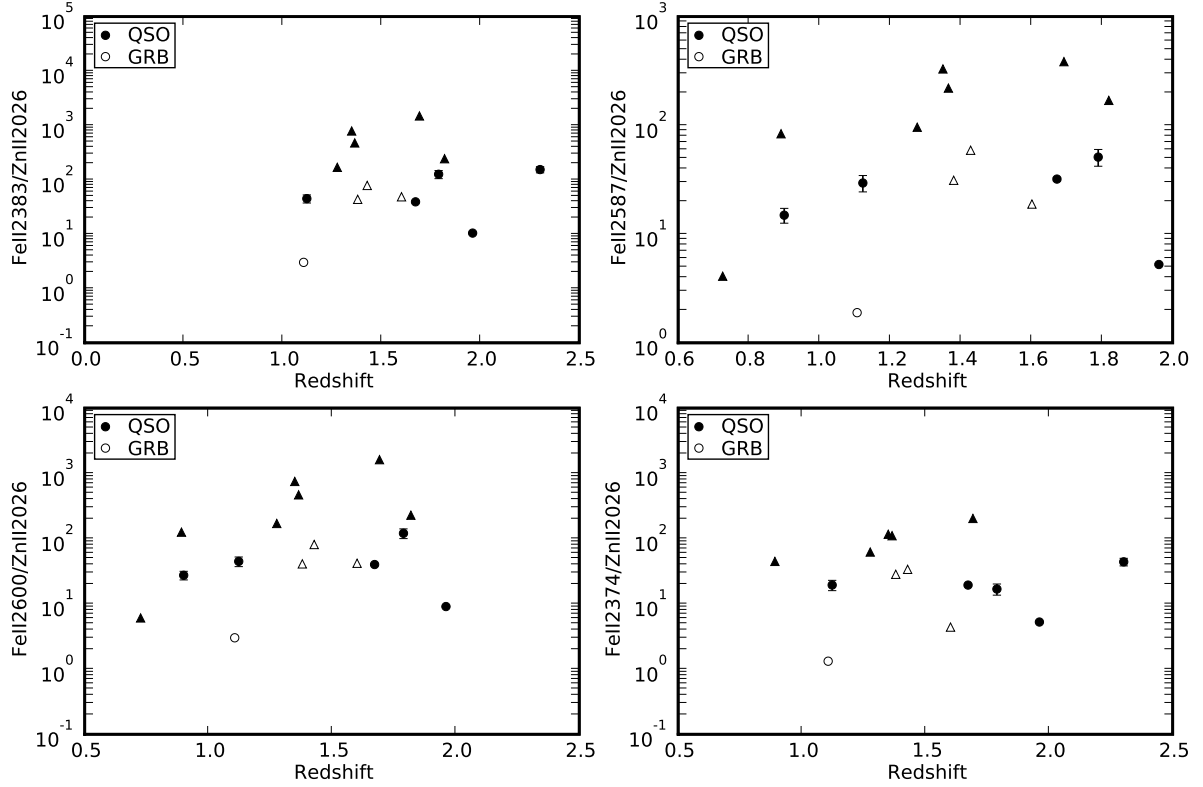}
\caption{\FeII/\ZnII ~equivalent width ratios for different
  transitions, as observed in GRB (open symbol) and quasar (filled
  symbol) absorption systems from our study.  Error bars are typically
  smaller than the symbol size; lower limits are shown as triangles.
  Low values of the various \FeII/\ZnII ~ratios indicate high
  depletion of gas onto dust grains.  Comparison of the two
  samples does not suggest any particular enrichment of dusty
  absorbers among the GRB absorber population, however, the most
  extreme (low-ratio) detection at $z \sim 1.1$ is for one of the
  systems found toward GRB\,060418 \citep{evl06}.  The distribution
  of ratio values, including lower limits, demonstrate that there is
  not an excess of dusty absorbers sufficient to bias
  magnitude-limited quasar surveys and thereby explain the $dN/dz$
  discrepancy between GRB and quasar lines of sight. }
\label{fig:fig8}
\end{figure*}
 \clearpage


\begin{figure*}
\figurenum{9}
\epsscale{0.9}
\plotone{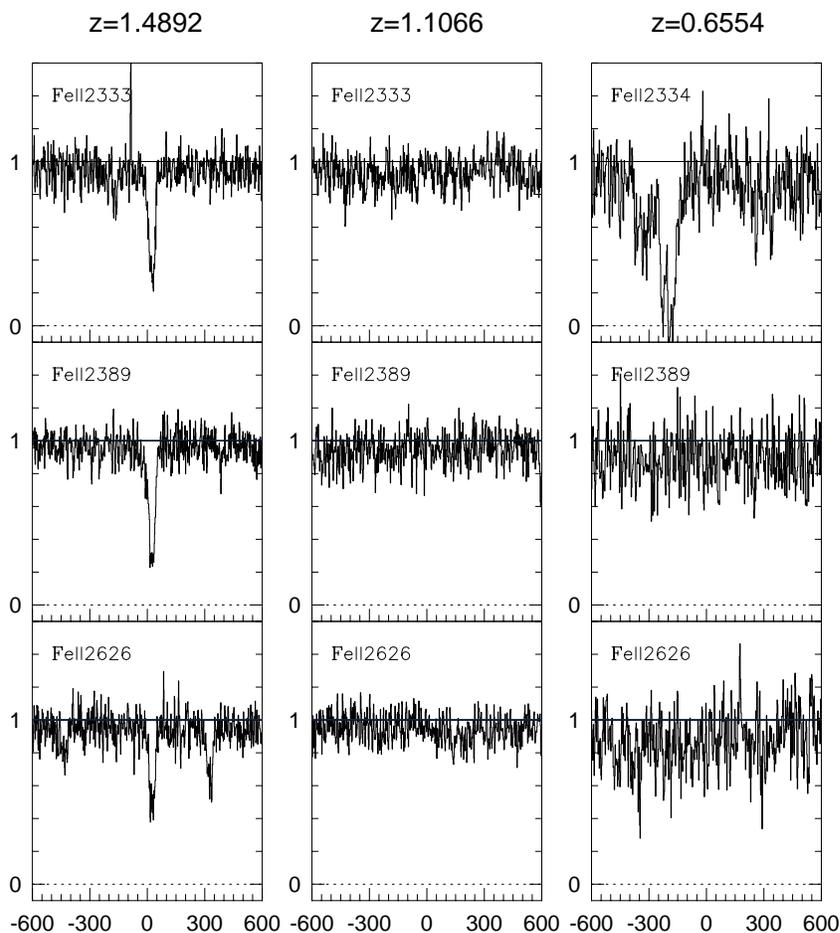}
\caption{\footnotesize{Comparison between GRB060418 {\FeII}* and {\NiII}*
  transitions at the host galaxy redshift ($z = 1.489$) and at the
  redhsifts of two intervening absorbers (1.106, 0.655). The continuum
  level is shown; detected features are identified near velocity zero,
  as derived from several other transitions. These absorption features
  are present in the host galaxy absorber but not in the intervening
  systems, consistent with a non-intrinsic (intervening galaxy) origin
  for these systems. The variability of these fine structure
  transitions, due to UV pumping from the GRB radiation field, places
  constraints on the location of the host galaxy absorber relative to
  the GRB \citep{vls07}. No other example of such fine-structure
  transitions has been seen in this or other GRB absorbers of our
  sample.}}
\label{fig:fig9}
\end{figure*}
\clearpage


\bibliographystyle{apj}
\bibliography{journals_apj,grboldbi,grbqso}

\end{document}